\documentclass[12pt]{article}
\usepackage{amssymb}

\newcommand{\C}{\mathbb{C}}
\newcommand{\R}{\mathbb{R}}
\newcommand{\GG}{{\mathcal{G}}}
\newcommand{\UU}{{\mathcal{U}}}
\newcommand{\QED}{\mbox{\rule[-1.5pt]{6pt}{10pt}}}
\newcommand{\restr}{\vert\hskip -5.5pt
   \phantom{\vert}^{\scriptscriptstyle\backslash}}
\newtheorem{claim}{Claim}[section]
\newtheorem{theorem}[claim]{Theorem}
\newtheorem{proposition}[claim]{Proposition}
\newtheorem{lemma}[claim]{Lemma}
\newtheorem{example}[claim]{Example}

\newtheorem{remark}[claim]{Remark}
\newtheorem{remarks}[claim]{Remarks}

\begin{document}

\title{A Model of Interband Radiative Transition}
\author{J.~Dittrich,$^{a,b}$ P.~Exner,$^{a,b}$ and M.~Hirokawa$^{c}$}
\date{}
\maketitle
\begin{quote}
{\small \em a) Department of Theoretical Physics, Nuclear Physics
Institute,\\ \phantom{e)x}Academy of Sciences, 25068 \v Re\v z,
Czech Republic \\
 b) Doppler Institute, Czech Technical University, B\v{r}ehov{\'a}
 7, \\ \phantom{e)x}11519 Prague, Czech Republic \\
 c) Department of Mathematics, Faculty of Science, Okayama \\
 \phantom{e)x}University, 3-1-1 Tsushima-naka, Okayama 700-8530, Japan \\
 \rm \phantom{e)x}dittrich@ujf.cas.cz, exner@ujf.cas.cz, \\
 \rm \phantom{e)x}hirokawa@math.okayama-u.ac.jp} \vspace{8mm}

\noindent {\small We consider a simple model which is a caricature
of a crystal interacting with a radiation field. The model has two
bands of continuous spectrum and the particle can pass from the
upper one to the lower by radiating a photon, the coupling between
the excited and deexcited states being of a Friedrichs type. Under
suitable regularity and analyticity assumptions we find the
continued resolvent and show that for weak enough coupling it has
a curve-type singularity in the lower halfplane which is a
deformation of the upper-band spectral cut. We then find a formula
for the decay amplitude and show that for a fixed energy it is
approximately exponential at intermediate times, while the tail
has a power-like behaviour.}
\end{quote}
\newpage


\renewcommand{\thesection}{\Roman{section}}
\section{Introduction}
\renewcommand{\thesection}{\arabic{section}}

A rigorous description of decay and resonance processes in quantum
theory has a long history starting from the Friedrichs model
presented in \cite{Friedrichs} and discussed later in numerous
papers -- see, e.g. \cite{AMKG}, \cite{DE}. A systematic study of
the problem started in the seventies. J.~Howland and
H.~Baumg\"artel with collaborators -- see \cite{Ho1}, \cite{Ho2},
\cite{BD}, \cite{BDW} and the papers quoted there -- used operator
methods to establish the existence of resonance poles and to prove
the Fermi rule for various systems with perturbed embedded
eigenvalues. At the same time the seminal paper \cite{AG} by
J.~Aguilar and J.-M.~Combes initiated the development of
complex-scaling methods which are nowadays a very efficient tool
to study resonances of Schr\"odinger operators.

In the eighties many papers dealing with quantum-field decay
models appeared. A phenomenological models based on the Langevin
equation were investigated in \cite{GWT}, \cite{APSJ}, \cite{JIG}
and \cite{BM}. Moreover a generalization of them was given by
A.~Arai \cite{Ar3} within the Hamiltonian formalism. In a last few
years the long-time behavior of canonical correlation functions
for general Hamiltonians was investigated in \cite{Hi} by applying
the results of \cite{Ar3} and \cite{Ar2} via a quantum Langevin
equation. From the point of view of virtual transitions, the
long-time behavior of a correlation function was studied in
\cite{KPPP}. It is also worth of noticing that, revisiting the
decay problem, Bach, Fr\"{o}hlich, and Sigal have developed a new
manner to analyze the resonace problems for a class of models in
quantum electrodynamics \cite{BFS1}, \cite{BFS2}.

In most of these models the unstable states come from perturbation
of eigenvalues, either embedded in the continuous spectrum or
isolated as in the case of Stark effect. Much less attention has
been paid to the situation when the states which should decay
belong to the continuous spectrum of the unperturbed Hamiltonian.
An archetypal example of such a situation is a crystal in which an
electron can radiate a photon and pass to a lower spectral band. A
natural model in this case would be a Schr\"odinger operator with
a periodic potential coupled to a quantized field. This is not
easy, however. To start with a simpler case, we discuss in this
paper a model of Friedrichs type with transitions between two
bands of the absolutely continuous spectra which can be regarded
as a one-photon approximation of the more realistic description.

While perturbed embedded eigenvalues typically give rise to resonance
poles in the analytically continued resolvent, we are going to show that
in the mentioned model the cut-like singularity corresponding to the
``excited'' spectral band gets deformed to the lower complex halfplane.
Recall that a similar behavior has been observed in a completely
different type of systems which involve a perturbation of a band
spectrum, namely for scattering in finitely periodic systems \cite{BG}.
Here we have a situation with a finite number of resonances which
accumulate, however, along curves in the lower halfplane which are close
to the spectral bands of the infinite system when the interaction is weak.

Let us describe briefly the contents of the paper. After
formulating the model in the next section we shall compute in
Section~III the projection of the Hamiltonian resolvent onto the
subspace of excited states corresponding to the upper spectral
band of the ``crystal''. Under natural regularity assumptions we
prove the mentioned claim about the change of the spectral
singularity caused by a decay with the radiation of a ``photon''.

Then we turn to the time evolution of the undecayed state and show
that its projection onto the upper-band subspace is -- at least
for a weak enough coupling -- realized as multiplication by a
function which we evaluate explicitly. The rest of the paper is
devoted to properties of this decay amplitude. We show that in the
weak-coupling case the latter is dominated at intermediate times
by an exponential function. Hence the population of the excited
spectral band changes in the course of the evolution: the
wavefunction components supported in the regions where the
deformed singularity is closer to the real axis survive longer. On
the other hand, similarly to the usual decay theory, the
deexcitation process cannot be purely exponential; we show that
the decay amplitude has a power-like tail at long times.


\renewcommand{\thesection}{\Roman{section}}
\section{Description of the model}
\renewcommand{\thesection}{\arabic{section}}

The ``crystal part'' of our model is assumed to have the simplest
nontrivial spectrum consisting of a pair of disjoint absolutely
continuous bands $I_0=[\xi_0^{(-)},\xi_0^{(+)}]$ and
$I_1=[\xi_1^{(-)},\xi_1^{(+)}]$ with
$-\infty<\xi_0^{(-)}<\xi_0^{(+)}<\xi_1^{(-)}<\xi_1^{(+)}<\infty$.
Using the spectral representation \cite{RS1} we can assume without
loss of generality that the crystal state space is $L^2(I_1\cup
I_0,w(x)\, dx)$ with the Hamiltonian $H_c$ acting as
multiplication by the variable $x$; the weight function $w$ is
positive a.e., Lebesgue integrable, and satisfies
 $$ \int_{I_1\cup I_0} w(x)\, dx=1 \,. $$
As we have said the ``field part'' is represented by the vacuum
and one-photon (or phonon) states, which coexist with the upper
and lower band of the ``crystal'', respectively. The photon vacuum
is by assumption a single state of zero energy, while the
single-photon states belong to the space
$L^2([\nu,\infty),\omega(z)\, dz)$, $\nu\geq 0$, on which the free
Hamiltonian $H_p$ acts as a multiplication by the variable $z$.
The weight function $\omega$ is again Lebesgue integrable,
non-negative a.e., and satisfies
 $$ \int_\nu^\infty\omega(z)\, dz=1 \,. $$
Putting the two components together we get the total state space
of our model in the form
\begin{equation}
{\mathcal{H}}= {\mathcal{H}}_0 \oplus {\mathcal{H}}_1 := L^2(I_1,w_1(x)\,
dx)\oplus [L^2(I_0,w_0(y)\, dy)\otimes L^2(K,\omega(z)\, dz)]\,,
\label{totHS}
\end{equation}
where $K=[\nu,\infty)$ and $w_{\alpha}:= w\restr I_\alpha,\,
\alpha=0,1$. The free Hamiltonian acts as
$$ H_0\left(\begin{array}{c} f\\ g\end{array}\right) =
\left(\begin{array}{c} H_cf\\ (\overline{H_c\otimes I + I\otimes
H_p})g\end{array}\right) $$
which means
\begin{equation} \label{H_0} \left( H_0\left(\begin{array}{c} f\\
g\end{array}\right) \right) \left(\begin{array}{c} x\\
y,z\end{array}\right) = \left(\begin{array}{c} xf(x)\\
(y+z)g(y,z)\end{array}\right) \end{equation}
with the arguments $x\in I_1$, $y\in I_0$, and $z\in K$.

Next we have to choose the interaction part of the Hamiltonian.
Being inspired by the Friedrichs model we require
   \begin{description}
   \vspace{-1.2ex}
   \item{\em (i)\phantom{i}} the interaction includes necessarily a single photon
   emission/ab\-sorp\-tion, or in other words, the projections of
   $H_{\mathrm{int}}$ on $L^2(I_1,w_1(x)\, dx)$ and its orthogonal
   complement in $\mathcal{H}$ are zero,
   \vspace{-1.2ex}
   \item{\em (ii)} the interaction is ``minimal'' in the sense
   that the action of $H_{\mathrm{int}}$ can be written in terms of
   multiplication by a ''formfactor'', integration, and possibly a
   change of variables.
   \vspace{-1.2ex}
   \end{description}
It follows from {\em (i)} that $H_{\mathrm{int}} = \kappa L$ with an
interaction constant $\kappa$ and an
``off-diagonal'' operator $L$, where $L_{ij}: {\mathcal{H}}_j \to
{\mathcal{H}}_i$, i.e.
\begin{equation} \label{H_int}
\left( H_{\mathrm{int}}\left(\begin{array}{c} f\\
g\end{array}\right) \right) \left(\begin{array}{c} x\\
y,z\end{array}\right) = \kappa \left(\begin{array}{c}
(L_{01}g)(x)\\ (L_{10}f)(y,z)\end{array}\right)\,.
\end{equation}
Furthermore, in accordance with {\em (ii)} the operator $L_{10}$
should be chosen in the form
\begin{equation}
(L_{10}f)(y,z)=\lambda(y,z)\,f(u(y,z))\,,
\end{equation}
where $\lambda: I_0\times K\to\C$ and $u: I_0\times K\to I_1$ are
functions containing the dynamical information about the system.
This choice in turn restricts $L_{01}$ because the full
Hamiltonian (with a real coupling constant $\kappa$) must be
symmmetric, which means
\begin{equation}
\int_{I_1}\overline{f(x)}(L_{01}g)(x)w_1(x)\,dx =
\int\!\!\int_{I_0\times K}
\overline{\lambda(y,z)f(u(y,z))}g(y,z)w_0(y)\omega(z)\,dy\,dz
\label{symgen}
\end{equation}
for all $f$ and $g$ from the operator domain. Suppose now that
there are functions $u,\,v$
such that $(y,z)\mapsto (u(y,z),v(y,z)): I_0\times K\to
I_1\times K$ is a bijective diffeomorphism which can be used as a
substitution at the r.h.s. of (\ref{symgen}) leading to
\begin{equation} (L_{01}g)(x)w_1(x) = \int_K
\overline{\lambda(y,z)} g(y,z)\left|\frac{D(y,z)}{D(u,v)}\right|
w_0(y)\omega(z)\,dt\,, \end{equation}
the variables $y,\,z$ being expressed as the inverse of
$x=u(y,z)$ and $t=v(y,z)$
at the r.h.s.
\begin{remarks}
{\rm (a) For the sake of simplicity, assume that $u$ depends on a
single variable mapping $I_0$ onto $I_1$. This will reduce the
dependence of the transition between a pair of states in $I_1$ and
$I_0$, respectively, on the photonic component of the system. \\
(b) In the same vein we could suppose that
\begin{equation}
\label{couplfactor} \lambda(y,z)=\lambda_0(y)\lambda_K(z)
\end{equation}
which will turn $H_0+ H_{\mathrm{int}}$ -- up to the isomorphism
between $I_1$ and $I_0$ -- into a direct integral of
Friedrichs-type Hamiltonians. However, we choose a nontrivial
setup and do not require that the dependence of the interaction
strength  on the energies of the excited state and the photon
contained in the function $\lambda$ factorizes. In other words, we
will keep a general $\lambda: I_0\times K\to\C$. } \end{remarks}

After this heuristic discussion, let us define the Hamiltonian
which we shall consider in the following. We suppose that
 \begin{description}
 \item{\em (a1)} $\;u:I_0\to I_1$ is a bijective $C^1$-diffeomorphism,
 \end{description}
then the interaction term $H_{\mathrm{int}}$ acts according to
(\ref{H_int}) with
\begin{eqnarray} (L_{10}f)(y,z) \!&:=&\! \lambda(y,z)f(u(y))\,,
\nonumber \\ (L_{01}g)(x) \!&:=&\!
\frac{w_0(u^{-1}(x))}{|u'(u^{-1}(x))|w_1(x)} \int_K
\overline{\lambda(u^{-1}(x),z)}g(u^{-1}(x),z)\omega(z)\, dz
\phantom{AAA} \end{eqnarray}
with $x\in I_1$, $y\in I_0$, and $z\in K$. The second expression
makes sense because the two factors in the denominator are
positive a.e. by assumption. The operator $L$ defined in this way
is formally symmetric and unbounded in general. To get a
self-adjoint Hamiltonian we add a boundedness assumption.
Specifically, we assume that
 \begin{description}
 \item{\em (a2)} $\lambda$ is Lebesgue measurable in
$I_0\times K$ and there are positive $C$, $C_1$ such that
$$ \int_K |\lambda(y,z)|^2\omega(z)\, dz \le C\,, \qquad w_0(y)
\le C_1\,|u'(y)|w_1(u(y)) $$
holds for every $y\in I_0$;
 \end{description}
the last inequality means that the Radon-Nikod\'{y}m derivative
appearing as the first factor in $L_{01}g$ is bounded.
 \begin{proposition}
Under the assumptions {\it (a1)} and {\it (a2)},
$H_{\mathrm{int}}$
is bounded and symmetric. Consequently,
$$ H=H(\kappa)= H_0+H_{\mathrm{int}}= H_0+\kappa L $$
is self-adjoint on the domain of $H_0$.
 \end{proposition}
{\sl Proof:} It remains to verify the boundedness of
$H_{\mathrm{int}}$ which amounts to checking that the operators
$L_{10}:\, {\mathcal{H}}_0\to {\mathcal{H}}_1$ and $L_{01}:\,
{\mathcal{H}}_1\to {\mathcal{H}}_0$ are bounded. This is easily seen
from the following estimates:
\begin{eqnarray*}
\|L_{10}f\|_{I_0\times K}^2 &=& \int\!\!\int_{I_0\times K}
|\lambda(y,z)|^2 |f(u(y))|^2 w_0(y)\omega(z)\, dy\, dz \\ &\leq& C
\int_{I_1} |f(x)|^2 \frac{w_0(u^{-1}(x))}{|u'(u^{-1}(x))|}\, dx
\,\le\, C_1C \| f\|_{I_1}^2
\end{eqnarray*}
and
\begin{eqnarray*}
\|L_{01}g\|_{I_1}^2 &=& \int_{I_1}
\left[\frac{w_0(u^{-1}(x))}{u'(u^{-1}(x))w_1(x)}\right]^2\,
\left[\int_K
{\lambda(u^{-1}(x),z)}\overline{g(u^{-1}(x),z)}\omega(z)\, dz
\right] \\ & & \times\left[ \int_K
\overline{\lambda(u^{-1}(x),t)}g(u^{-1}(x),t)\omega(t)\, dt
\right] w_1(x)\, dx \\ &=&
\int_{I_0}\,\frac{w_0(y)^2}{|u'(y)|w_1(u(y))}
\\ & & \times\left\{\int\!\! \int_{K\times K}
\left[\lambda(y,z)\overline{\lambda(y,t)}\right]\,
\left[\overline{g(y,z)}g(y,t)\right]\omega(z)\omega(t)\,dz\,dt
\right\} \,dy \\ &\leq& C_1 \int_{I_0}
\left[\int_K|\lambda(y,z)|^2\omega(z)\,dz\right]\,
\left[\int_K|g(y,t)|^2\omega(t)\,dt\right]w_0(y)\,dy \\ &\leq&
C_1C\|g\|_{I_0\times K}^2,
\end{eqnarray*}
where we have used the Fubini theorem in combination with the
Schwarz inequality for the scalar product in $L^2(K\times
K,\omega(z)\omega(t)\,dz\,dt)$. \quad \QED \vspace{2mm}

Before proceeding further let us make a comment on the
assumptions, part physical and part technical, which we will have
to make in the following. Since the present model is rather a
motivation study for a more realistic one, we do not strive for
the maximal possible generality. On the other hand, we do not want
to impose many unnecessary restrictions which would correspond to
a fully specific system such as the one given below.
\begin{example} \label{1D crystal} {\rm Let $E_j(\cdot),\: j=0,1$
be the lowest two dispersion curves of a one-dimensional crystal.
Since we are discussing a caricature model, we neglect the
multiplicity of the eigenvalues. In other words, we consider just
a half of the Brillouin zone and regard $E_j$ as maps $[0,\pi]\to
I_j$ with $E_0$ strictly increasing and $E_1$ strictly decreasing.
Moreover, both are restrictions to $[0,\pi]$ of real-analytic
functions with the  first derivatives vanishing at the endpoints
of the interval but nonzero in its interior.

To rewrite the band projections of the crystal Hamiltonian in our
formalism, we employ the operators $U_j: L^2([0,\pi]) \to
L^2(I_j,w_j(y)\, dy)$ defined by $(U_j f)(y):= f(E_j^{-1}(y))$;
the definition makes sense since the inverse functions $E_j^{-1}$
exist by assumption. The operators $U_j$ are unitary provided we
put
\begin{equation}
\label{1D w} w_j(y) = |E'_j(E_j^{-1}(y))|^{-1}\,.
\end{equation}
These functions are $C^{\infty}$ in $(0,\pi)$ with singularities
at the endpoints but the latter are integrable. In particular, if
$E''_j(\vartheta)\ne 0$ at $\vartheta=0,\pi$ we have $w_j(y) =
O\left(|y\!-\! \xi_j^{(\pm)}|^{-1/2}\right)$ there.

One of the basic ingredients is, of course, the function $u$.
Since the system of the crystal plus the radiation field is
invariant w.r.t. the discrete group of translations on a multiple
of the lattice constant, it is natural in the present example to
suppose that the interaction does not couple states whose
quasimomentum support in the upper and lower bands are disjoint.
This is achieved if we choose
\begin{equation}
\label{1D u} u(y) = E_1\left(E_0^{-1}(y)\right) \;;
\end{equation}
it is easy to see that it is a $C^{\infty}$ function and
\begin{equation}
\label{1D u'} u'(y) = {E'_1\left(E_0^{-1}(y)\right) \over
E'_0\left(E_0^{-1}(y)\right)}
\end{equation}
has finite limits at $\xi_0^{(\pm)}$ assuming that $E_0$ and $E_1$
have the first non-vanishing derivative at $0$ resp. $\pi$ of the
same order. On the other hand we think of the radiation field as
of the electromagnetic field in the rotating wave approximation.
In this case we put the threshold energy $\nu=0$ and $\omega(z) =
\chi_{[0,\nu_{\mathrm{max}}]}(z)$ where $\nu_{\mathrm{max}}$ is a
possible ultraviolet cut-off.

Under these model assumptions {\em (a1)} is satisfied
automatically and the same is true for the second part of {\em
(a2)}; it follows from (\ref{1D w}) and (\ref{1D u'}) that it is
valid for any $C_1\ge 1$. The only remaining restriction is thus
the boundedness condition $\int_0^{\nu_{\mathrm{max}}} |\lambda
(y,z)|^2 \,dz \le C$ for the formfactor. }
\end{example}


\setcounter{equation}{0}
\renewcommand{\thesection}{\Roman{section}}
\section{The resolvent}
\renewcommand{\thesection}{\arabic{section}}

As usual the spectral information is contained in the resolvent of
the Hamiltonian. Under our assumptions, we can find it explicitly
by solving the equation
$$ (H-\zeta)\left(\begin{array}{c} f\\ g\end{array}\right) =
\left( \begin{array}{c} f_1\\ g_1\end{array}\right) $$
for $\zeta$ in the resolvent set, in particular, for all
$\zeta\in\C\setminus\R$. It is straightforward to check that
\begin{eqnarray} \label{resolv} f(x) \!&=&\! r(x,\zeta)f_1(x) -
\kappa r(x,\zeta)\,\frac{w_0(u^{-1}(x))}{|u'(u^{-1}(x))|w_1(x)}
\nonumber \\ && \times
\int_K\frac{\overline{\lambda(u^{-1}(x),z)}} {u^{-1}(x)+z-\zeta}
\, g_1(u^{-1}(x),z)\, \omega(z)\, dz \,, \\ g(y,z) \!&=&\!
-\,\kappa\frac{\lambda(y,z)}{y+z-\zeta}\, r(u(y),\zeta)\,
f_1(u(y)) \, +\,\frac{g_1(y,z)}{y+z-\zeta} +
\kappa^2\frac{\lambda(y,z)}{y+z-\zeta} \nonumber \\ && \times\,
r(u(y),\zeta)\, \frac{w_0(y)}{|u'(y)|w_1(u(y))}
\int_K\frac{\overline{\lambda(y,r)}}{y+r-\zeta}\, g_1(y,r)\,
\omega(r)\, dr\,, \nonumber \end{eqnarray}
where
$$ r(x,\zeta):=\left\{ x - \zeta - \kappa^2
\frac{w_0(u^{-1}(x))}{|u'(u^{-1}(x))|w_1(x)}
\int_K\frac{|\lambda(u^{-1}(x),z)|^2}{u^{-1}(x)+z-\zeta}\,
\omega(z)\, dz\right\}^{-1} \; . $$
Let $P$ be the projection onto the subspace ${\mathcal{H}}_0=
L^2(I_1,w_1\, dx)$ of ``undecayed'' states in $\mathcal{H}$,
$$ P\left(\begin{array}{c} f_1\\ g_1\end{array}\right) =
\left(\begin{array}{c} f_1\\ 0\end{array}\right)\,. $$
By (\ref{resolv}), the reduced resolvent acts then as
multiplication by the function $r$,
\begin{equation}
\label{rresolv} P(H-\zeta)^{-1}P=r(\cdot,\zeta) P\,.
\end{equation}
For the sake of brevity we introduce the following notation,
\begin{eqnarray}
v(y,z)&:=&|\lambda(y,z)|^2\omega(z) \,, \\ \varrho(x) &:=&
\frac{w_0(u^{-1}(x))}{|u'(u^{-1}(x))| w_1(x)} \,, \\ \label{fG}
\GG(y,\zeta) &:=& \int_K\frac{v(y,z)}{y+z-\zeta}\, dz\,,
\end{eqnarray}
so the function $r$ can be written as
\begin{equation}
\label{rresolvform}
r(x,\zeta)=\left\{ x - \zeta - \kappa^2
\varrho(x)\,\GG(u^{-1}(x),\zeta)\right\}^{-1}
\end{equation}
for $\Im \zeta \not= 0$.
\begin{remark}
{\rm In the particular case of Example~\ref{1D crystal} it follows
from (\ref{1D w}) and (\ref{1D u'}) that $\varrho(x)=1$, and
moreover, $v(y,z) = |\lambda (y,z)|^2
\chi_{[0,\nu_{\mathrm{max}}]}(z)$. }
\end{remark}
To reveal the analytic properties of $r(x,\cdot)$ let us begin
with those of $\GG(y,\cdot)$.
\begin{lemma}
\label{principal} Let $v(y,\cdot)$ have a locally bounded
derivative in $(\nu,\infty)$. Then for any $y\in I_0$ and a real
$\zeta>y+\nu$ there exists finite principal value of the integral
\begin{equation}
I(y,\zeta):= {\mathcal{P}}
\int_\nu^{\infty}\frac{v(y,z)}{y+z-\zeta}\, dz \,.
\end{equation}
Moreover, for any $k\in (0,{\zeta-y-\nu})$,
\begin{eqnarray}
\nonumber
I(y,\zeta)&=&\int_\nu^{\zeta-y-k}\frac{v(y,z)}{y+z-\zeta}\, dz +
\int_{\zeta-y-k}^{\zeta-y+k}\frac{v(y,z)-v(y,\zeta-y)}{y+z-\zeta}\,
dz\\ & & +\int_{\zeta-y+k}^{\infty}\frac{v(y,z)}{y+z-\zeta}\, dz
\label{pvconv}
\end{eqnarray}
where all the three integrals are Lebesgue convergent.
\end{lemma}
{\sl Proof:} Choose any $k\in (0,{\zeta-y-\nu})$. As the
integrals
$$ \int_\nu^{\zeta-y-k}\frac{v(y,z)}{y+z-\zeta}\, dz \quad
\mathrm{and} \quad
\int_{\zeta-y+k}^\infty\frac{v(y,z)}{y+z-\zeta}\, dz $$
exist due to the assumption {\em (a2)} it is sufficient to check
the convergence of
\begin{equation}
I_k(y,\zeta)={\cal
P}\int_{\zeta-y-k}^{\zeta-y+k}\frac{v(y,z)}{y+z-\zeta}\, dz \; .
\end{equation}
We employ the identity $v(y,z) =v(y,\zeta-y)
+[v(y,z)-v(y,\zeta-y)]$ together with the estimate
$$ |v(y,z)-v(y,\zeta-y)|\leq c_1 |y+z-\zeta| $$
with a finite $c_1$ independent of $z$. We see that finite
$$ \int_{\zeta-y-k}^{\zeta-y+k}
\frac{v(y,z)-v(y,\zeta-y)}{y+z-\zeta}\, dz $$
exists and it is sufficient to check ${\cal P}
\int_{\zeta-y-k}^{\zeta-y+k} \frac{dz}{y+z-\zeta}$ which is easily
seen to exist and to be equal to zero. \quad \QED \vspace{3mm}

As usual in similar situations to proceed one needs some
analyticity assumption about the formfactor. In the present case
we suppose that
 \begin{description}
 \item{\em (a3)} for all $y\in I_0$ the function
$v(y,\cdot)$ can be holomorphically extended to an open set
$\Omega_{v,y}\supset (\nu,\infty)$; we denote the extension again
as $v(y,\cdot)$. Let us further assume that there is an open set
$\Omega$ in $\C$ such that
$$ (\xi_0^{(-)}+\nu,\infty)\subset\Omega\subset\cap_{y\in
I_0}(y+\Omega_{v,y}) \quad. $$
 \end{description}
Notice that the hypothesis of the previous lemma is satisfied
under {\em (a3)}. Now we can make the following claim.
\begin{lemma}
\label{uplimlemma} Let $y\in I_0$ and $\xi\in (y+\nu,\infty)$.
Then
$$ \label{uplim} \lim_{\pm\Im \zeta>0,\,
\zeta\to\xi}\GG(y,\zeta)=I(y,\xi) \pm i\pi v(y,\xi-y)\,. $$
\end{lemma}
{\sl Proof:} Let us write again $\GG(y,\zeta)$ defined by
(\ref{fG}) as a sum of three integrals over the intervals
$(\nu,\xi-y-k)$, $(\xi-y-k,\xi-y+k)$ and
$(\xi-y+k,\infty)$ with $0<k<{\xi-y-\nu}$. The
first and the third integral can be interchanged with limit by
dominated convergence. The set $\Omega_{v,y}$ is open and contains
$(\nu,\infty)$, hence there is $k_1>0$ such that any
$\vartheta\in\C$
satisfying $|\vartheta-\xi+y|\leq k_1$ belongs to $\Omega_{v,y}$. Let us
consider only $\zeta$ satisfying $|\zeta-\xi|\leq k_1$
(so that $\zeta-y\in\Omega_{v,y}$)
in the second integral and denote
$\zeta_1=\Re \zeta$, then we employ the identity $v(y,z)
=v(y,\zeta_1-y)+[v(y,z)-v(y,\zeta_1-y)]$ and observe that
$$ |v(y,z)-v(y,\zeta_1-y)|\leq c_1(y,\xi,k,k_1)\, |y+z-\zeta_1|
\,. $$
The contribution from the difference can be thus also handled by
dominated convergence. In view of (\ref{pvconv}) we get
$$ \lim_{\pm\Im \zeta>0,\, \zeta\to\xi}\GG(y,\zeta) = I(y,\xi) +
v(y,\xi-y)\,\lim_{\pm\Im \zeta>0,\, \zeta\to\xi}
\int_{\xi-y-k}^{\xi-y+k}\frac{dz}{y+z-\zeta} $$
and the result follows by an easy calculation. \quad \QED

\begin{lemma}
\label{analcont} Define the functions $\GG_\Omega: I_0\times\Omega
\to\C$ and $\GG^\Omega: I_0\times\Omega \to\C$ by
\begin{equation}
\label{GGOdef}
\GG_\Omega(y,\zeta)=\left\{
\begin{array}{lcr}
\GG(y,\zeta)& \quad\dots\quad &\Im \zeta>0\\ I(y,\zeta)+i\pi
v(y,\zeta-y)& \quad\dots\quad &\Im \zeta=0\\ \GG(y,\zeta)+2i\pi
v(y,\zeta-y)& \quad\dots\quad &\Im \zeta<0
\end{array}
\right.
\end{equation}
\begin{equation}
\label{GGOdef2}
\GG^\Omega(y,\zeta)=\left\{
\begin{array}{lcr}
\GG(y,\zeta)-2i\pi
v(y,\zeta-y)& \quad\dots\quad &\Im \zeta>0\\
I(y,\zeta)-i\pi
v(y,\zeta-y)& \quad\dots\quad &\Im \zeta=0\\
\GG(y,\zeta)& \quad\dots\quad &\Im \zeta<0
\end{array}
\right.
\end{equation}
Under our assumptions {\em (a1)}--{\em (a3)}, the functions
$\GG_\Omega(y,\cdot)$ and $\GG^\Omega(y,\cdot)$ are holomorphic in
$\Omega\setminus(-\infty,y+\nu]$ for any fixed $y\in I_0$.
\end{lemma}
{\sl Proof:} By Lemma \ref{principal} and assumption {\em (a3)},
$\GG_\Omega$ is a finite function. Notice that
$\zeta-y\in\Omega_{v,y}$ for $\zeta\in\Omega$ and $y\in I_0$.
According to Lemma \ref{uplimlemma}, the function
$\GG_\Omega(y,\cdot)$ is continuous in $\{\zeta\in\Omega|\Im
\zeta\geq 0\}\setminus(-\infty,y+\nu]$  -- see, e.g., Thm 146 in
Ref. \cite{JarDII}.
Alternatively, the continuity of $I(y,\cdot)$ in
$(y+\nu,\infty)$ can be established directly from the dominated
convergence used in the proof of Lemma \ref{principal}.
Similarly, the continuity in
$\{\zeta\in\Omega\,|\, \Im\zeta\leq 0\}\setminus(-\infty,y+\nu]$
is seen and thus  $\GG_\Omega(y,\cdot)$ is
continuous in $\Omega\setminus(-\infty,y+\nu]$. As it is holomorphic in
$\{\zeta\in\Omega|\,\Im \zeta>0\}\cup\{\zeta\in\Omega|\,\Im
\zeta<0\}$ it is also holomorphic in $\Omega\setminus(-\infty,y+\nu]$
due to a corollary
(dubbed the {\em edge-of-wedge} theorem) of the Morera's
theorem (stating that the continuous function is holomorphic iff
the integrals over all rectangles with the edges parallel to the
axes are zero -- see, e.g., \cite[Thm~168]{Cerny} or
\cite[Thm~10.17]{Rudin}).
As to $\GG^{\Omega}(y , \cdot)$, we can prove
our statement in the same way as for $\GG_{\Omega}(y , \cdot)$.
\quad \QED \vspace{2mm}

Now we are in position to show what happens with the upper spectral
band under influence of the perturbation. Let us formulate some
further assumptions before.
\begin{description}
\item{\em (a4)}
The functions $\varrho(x)\GG_\Omega(u^{-1}(x),\zeta)$ and
$\varrho(x)\frac{\partial \GG_\Omega(u^{-1}(x),\zeta)}{\partial
\zeta}$ are continuous in the set $\{(x,\zeta)\in I_1\times \Omega
| \zeta\not\in (-\infty, u^{-1}(x)+\nu] \}$.
\item{\em (a5)}
For all $x\in I_1$, $$ x>u^{-1}(x)+\nu \,.$$
\end{description}
\begin{remarks}
{\rm (a) In the particular case of Example~\ref{1D crystal} the
factor $\varrho(x)=1$ can be dropped in {\em (a4)} and the
assumption {\em (a5)} is satisfied. \\ (b) While most assumptions
we make are of a technical nature, {\em (a5)} is a physical
hypothesis saying that in no part of the excited spectral band the
decay is prevented by energy conservation. It is satisfied, of
course, if $\nu=0$. }
\end{remarks}
Let us denote $$ \Omega_x=\Omega\setminus (-\infty,
u^{-1}(x)\!+\!\nu\,] \,. $$
\begin{theorem}
\label{singsimple} Assume (a1)--(a5). Then the following statements hold.\\
(a)
There exist $\Delta>0$, $\delta>0$ and a unique function $\zeta:
I_1\times (-\delta,\delta) \to \C$ satisfying
\begin{eqnarray}
\label{resingint} \zeta(x,\kappa) \in
(x-\Delta,x+\Delta)+i(-\Delta,\Delta) \subset\Omega_x\,,\\
\label{resing} x - \zeta(x,\kappa) - \kappa^2
\varrho(x)\,\GG_\Omega(u^{-1}(x),\zeta(x,\kappa))=0 \,.
\end{eqnarray}
The function $\zeta$ is continuous in $I_1\times (-\delta,\delta)$
and $\zeta(x,\cdot)\in C^{\infty}(-\delta,\delta)$. \\ (b) The
resolvent has no singularity in the upper complex half-plane,
in particular
\begin{equation} \label{singnonpos} \Im \zeta(x,\kappa)\leq 0
\end{equation}
holds for all $x\in I_1$, $\kappa\in (-\delta,\delta)$. Moreover,
if
\begin{equation}
\label{vnonzero} \varrho(x)v(u^{-1}(x),x-u^{-1}(x))\not= 0
\end{equation}
for all $x$ from some compact $I'\subset I_1$, then there exists a
$\delta\in(0,\delta_1]$ such that
\begin{equation}
\label{singnegative} \Im \zeta(x,\kappa)<0
\end{equation}
holds for all $0<|\kappa|<\delta_1$ and $x\in I'$.
\end{theorem}
{\sl Proof:} (a)  Let us denote
\begin{equation} \label{implf}
D_{+}(x,\kappa,\zeta) := x - \zeta - \kappa^2
\varrho(x)\,\GG_\Omega(u^{-1}(x),\zeta) \,.
\label{eq:D+}
\end{equation}
The functions $D_{+}$ and $\frac{\partial D_{+}}{\partial \zeta}$
are continuous in $\{(x,\kappa,\zeta)| x\in I_1, \kappa\in\R,
\zeta\in\Omega_x \}$ by assumption and $D_{+}(x,\cdot,\cdot)\in
C^\infty(\R\times\Omega_x)$ by Lemma \ref{analcont}. Furthermore,
$D_{+}(x,0,x)=0$ and
$$ \frac{\partial
D_{+}(x,0,x)}{\partial \zeta} = -1 \not= 0\, . $$
By the implicit function theorem  -- see, e.g.
\cite[Thm~211]{JarDII} -- to any $x_0\in I_1$ there exist
$d_{x_0}>0$, $\delta_{x_0}>0$ and $\Delta_{x_0}>0$ such that for
all $x\in(x_0-d_{x_0},x_0+d_{x_0})\cap I_1$ and $\kappa\in
(-\delta_{x_0},\delta_{x_0})$ there is just one
$\zeta_{x_0}(x,\kappa)\in (x_0-\Delta_{x_0},
x_0+\Delta_{x_0})+i(-\Delta_{x_0},\Delta_{x_0})
 \subset\Omega_x$
(recall {\em (a5)}) satisfying
$D_{+}(x,\kappa,\zeta_{x_0}(x,\kappa))=0$, i.e. the relation
(\ref{resing}). The function $\zeta_{x_0}$ is continuous in
$((x_0-d_{x_0},x_0 +d_{x_0})\cap
I_1)\times(-\delta_{x_0},\delta_{x_0})$ and
$\zeta_{x_0}(x,\cdot)\in C^\infty(-\delta_{x_0},\delta_{x_0})$ for
any fixed $x\in (x_0-d_{x_0},x_0+d_{x_0})$. We put
$$ d'_{x_0}=\min(\Delta_{x_0},d_{x_0})\, . $$
As $I_1$ is compact by assumption, the open covering of $I_1$
defined in this way has a finite subcovering, i.e. there exist a
finite number of points $x_j\in I_1$, $j=1,\dots,n$, such that
$$ I_1\subset\cup_{j=1}^n K_j\,; $$
we employ here the notation
$$ K_j=(x_j-d'_{x_j},x_j+d'_{x_j})\,,\quad
J_j=K_j+i(-d'_{x_j},d'_{x_j}) $$
for $j=1,\dots,n$. Let us pick a point $x_{jk}\in K_j\cap K_k$
for given $j,k=1,\dots,n$; then there is $0<\delta_{jk}\leq
\min(\delta_{x_j},\delta_{x_k})$ such that
\begin{equation}
\zeta_{x_j}(x_{jk},\kappa)\in J_j\cap J_k\,, \quad
\zeta_{x_k}(x_{jk},\kappa)\in J_j\cap J_k
\end{equation}
for $|\kappa|<\delta_{jk}$. Moreover, $\zeta_{x_j}(x,\kappa)
=\zeta_{x_k}(x,\kappa)$ for all $x\in K_j\cap K_k$ and
$|\kappa|<\delta_{jk}$; otherwise the uniqueness of $\zeta_{x_j}$
and $\zeta_{x_k}$ would be violated near at least one of the
points
$$ \sup \{x\in K_j\cap K_k\, | \, x\geq x_{jk},
\zeta_{x_j}(y,\kappa)=\zeta_{x_k}(y,\kappa)\, {\rm for}\,
x_{jk}\leq y\leq x, |\kappa|<\delta_{jk}\}\, , $$
$$ \inf \{x\in K_j\cap K_k\, | \, x\leq x_{jk},
\zeta_{x_j}(y,\kappa)=\zeta_{x_k}(y,\kappa)\, {\rm for}\, x\leq y\leq
x_{jk}, |\kappa|<\delta_{jk}\}\,. $$
Choosing a number $\delta'>0$ with $\delta'\leq\min_{1\leq j\leq
n}\delta_{x_j}$ and $\delta'\leq\min_{K_j\cap
K_k\not=\emptyset}\delta_{jk}$, we conclude that there exists a
unique $\zeta: I_1\times (-\delta',\delta')\to\C$ such that
$\zeta(x,\kappa)\in J_j$ for $x\in K_j$ and
$D_{+}(x,\kappa,\zeta(x,\kappa))=0$. The function $\zeta$ is
continuous in $I_1\times (-\delta',\delta')$ and
$\zeta(x,\cdot)\in C^\infty ((-\delta',\delta'))$ for any fixed
$x\in I_1$. Put
$$ h_j(x)=\min(x-x_j+d'_{x_j}, x_j+d'_{x_j}-x)\,. $$
The function $h_j:I_1\to\R$ defined in this way is continuous and
$x\in K_j$ if and only if $h_j(x)>0$. Then
$$ h(x):=\max_{1\leq j\leq n}h_j(x) $$
specifies a positive continuous function $h$ on $I_1$. Let us
denote
$$ D=\min_{x\in I_1}h(x) > 0\,,\quad
\Delta=\min\left(D,\min_{1\leq j\leq n}d'_{x_j}\right)
> 0\,. $$
As $\zeta$ is uniformly continuous on compact subsets of
$I_1\times (-\delta',\delta')$ there exists $0<\delta\leq\delta'$
such that
$$ \zeta(x,\kappa)\in (x-\Delta,x+\Delta) + i(-\Delta,\Delta) $$
for $x\in I_1$ and $|\kappa|<\delta$; hence the existence of the
numbers $\delta$, $\Delta$ and the function $\zeta$ is
demonstrated.

Finally, to check the uniqueness of $\zeta$ let us assume that
$\tilde{\zeta}$ is another function satisfying
$$ \tilde{\zeta}(x,\kappa)\in
(x-\Delta,x+\Delta)+i(-\Delta,\Delta)\,, \quad
D_{+}(x,\kappa,\tilde{\zeta}(x,\kappa))=0 $$
for $x\in I_1$, $\kappa\in (-\delta,\delta)$. Suppose that $x\in
I_1$ and $|\kappa|<\delta$ are given. There exists an index
$j=1,\dots,n$ such that
$$ h_j(x)=h(x)\geq D\,, \quad x\in K_j\,,\quad \zeta(x,\kappa)\in
J_j\,. $$
As the inequalities
$$ x-x_j+d'_{x_j}\geq D\,,\quad x_j+d'_{x_j}-x\geq D\,,\quad
-D<y-x<D $$
hold, where $y:=\Re\tilde{\zeta}(x,\kappa)$, we have also
$$ y-x_j+d'_{x_j}>0\,,\quad x_j+d'_{x_j}-y>0 $$
and $y\in K_j$. Furthermore,  $|\Im\tilde{\zeta}(x,\kappa)
|<\Delta\leq d'_{x_j}\leq\Delta_{x_j}$. Then $\zeta(x,\kappa)
=\tilde{\zeta}(x,\kappa)$ and the uniqueness is proven. \\
(b) Assume first that $\Im \zeta >0$, then
$\Im\GG_\Omega(u^{-1}(x),\zeta) =\Im G(u^{-1}(x),\zeta)\geq 0$ by Eqs.
(\ref{fG}) and (\ref{GGOdef}), so the r.h.s of Eq. (\ref{implf}) has
negative imaginary part. Consequently, there are no solutions
$\zeta(x,\kappa)$ with positive imaginary parts, in other words
(\ref{singnonpos}) holds.
We have checked here only that the open upper half-plane is a part of
the resolvent set for the Hamiltonian. In the lower half-plane, the
function $D_{+}(x,\kappa,\cdot)^{-1}$ is a meromorphic continuation of
$r(x,\cdot)$ and may have singularities.

Suppose now that (\ref{vnonzero}) holds. The expression
$\varrho(x) \frac{\partial \GG_\Omega(u^{-1}(x),
\zeta(x,\kappa))}{\partial \zeta}$ is continuous in
$(x,\kappa)\in I_1\times (-\delta,\delta)$. It follows that
$$ M:=\max_{(x,\kappa)\in I'\times
[-\frac{\delta}{2},\frac{\delta}{2}]} \left|\varrho(x)
\frac{\partial \GG_\Omega(u^{-1}(x),\zeta(x,\kappa))}{\partial
\zeta}\right|<\infty \,. $$
Differentiating the equation defining $\zeta(x,\kappa)$ with
respect to $\kappa^2$ we get
\begin{equation}
\label{kappader1} \frac{\partial
\zeta(x,\kappa)}{\partial(\kappa^2)} +
\varrho(x)\GG_\Omega(u^{-1}(x),\zeta(x,\kappa)) +
\kappa^2\varrho(x)\frac{\partial\GG_\Omega(u^{-1}(x),\zeta(x,\kappa))}
{\partial \zeta}\frac{\partial \zeta(x,\kappa)}{\partial
(\kappa^2)} = 0\,.
\end{equation}
In combination with the previous inequality we conclude that
\begin{equation}
\label{kappader2} \frac{\partial
\zeta(x,\kappa)}{\partial(\kappa^2)} =
-\frac{\varrho(x)\GG_\Omega(u^{-1}(x),\zeta(x,\kappa))} {1 +
\kappa^2\varrho(x) \frac{\partial
\GG_\Omega(u^{-1}(x),\zeta(x,\kappa))}{\partial \zeta} }
\end{equation}
is continuous in $(x,\kappa)\in I'\times
\left(-\min(\frac{\delta}{2}, M^{-\frac{1}{2}}),
\min(\frac{\delta}{2},M^{-\frac{1}{2}})\right)$ defining
$M^{-\frac{1}{2}}=\infty$ for $M=0$. Furthermore,
the assumption (\ref{vnonzero}) together with (\ref{GGOdef})
implies $\Im \frac{\partial \zeta(x,0)}{\partial(\kappa^2)} <
0$, hence there is
$0<\delta_1\leq\min(\frac{\delta}{2},M^{-\frac{1}{2}})$ such that
$$ \Im \frac{\partial \zeta(x,\kappa)}{\partial(\kappa^2)} < 0
$$
for $(x,\kappa)\in I'\times(0,\delta_1)$ and
(\ref{singnegative}) holds. \quad \QED \vspace{2mm}

\begin{remarks}
{\em (a) Putting $\kappa=0$ in (\ref{kappader1}) we obtain
\begin{equation}
\frac{\partial \zeta(x,0)}{\partial (\kappa^2)} =
-\varrho(x)\GG_\Omega(u^{-1}(x),x)\,,
\end{equation}
where right-hand side is given by Lemma \ref{analcont}. This
relation can be regarded as an analogue of the Fermi golden rule
in the present situation.
\\ [1mm]
(b) Notice that for the factorization (\ref{couplfactor}) the term
$|\lambda_0(u^{-1}(x))|^2$ factorizes from
$\GG_\Omega(u^{-1}(x),\zeta)$ and $\zeta(x,\kappa)=x$ holds
whenever $\lambda_0(u^{-1}(x))=0$.}
\end{remarks}

\setcounter{equation}{0}
\renewcommand{\thesection}{\Roman{section}}
\section{Decay of excited states}
\renewcommand{\thesection}{\arabic{section}}

In accordance with the physical motivation, we are interested in
transitions from a given state supported in $I_1$ into those
in $I_0$. To find the time profile of the de-excitation
probability it is sufficient to know  the reduced evolution
operator $PU(t)P=Pe^{-iHt}P$. Suppose that the initial state is
of the form
$$ \Psi_0=\left(\begin{array}{c} \psi_0\\ 0\end{array}\right)\, ,
$$
for some $\psi_0\in L^2(I_1,w_1(x)\,dx)$ with $\|\Psi_0\|^2=\int_{I_1}
|\psi_0(x)|^2\,w_1(x)\,dx=1$. Its time evolution is given by the
Stone formula,
$$ U(t)\Psi_0=\lim_{\eta\to 0^+}\frac{1}{2\pi i}\int_{\R}
\left[(H-\xi-i\eta)^{-1} - (H-\xi+i\eta)^{-1}\right] e^{-i\xi
t}\Psi_0\, d\xi\,, $$
according to \cite[Thm~VIII.5]{RS1}, and the projection $P$ can
be interchanged with the limit and the integral being a bounded
operator. This yields the reduced evolution operator,
$$ P U(t)\Psi_0 = \left(\begin{array}{c} \psi(t,\cdot)\\
0\end{array}\right)\,, $$
where
\begin{equation}
\label{redevol}
\psi(t,\cdot) = \lim_{\eta\to 0^+}\frac{1}{2\pi i}\int_{\R}
\left[r(\cdot,\xi+i\eta) - r(\cdot,\xi-i\eta)\right] e^{-i\xi t}
\psi_0(\cdot)\, d\xi
\end{equation}
and $r$ is given by (\ref{rresolvform}). The integral and the
limit refer to functions with values in ${\mathcal{H}}_0
=L^2(I_1,w_1(x)\, dx)$; they are known to be convergent as the
Hamiltonian $H$ is self-adjoint.

Let us now look for conditions under which the interchange of the
limit and the integral in (\ref{redevol}) is possible. To this
end, we need more assumptions.
\begin{description}
\item{\em (a6)} $\; v(y,z) \leq C_2\;$ and $\; \left|\frac{\partial
v(y,z)}{\partial z}\right|\leq C_3\;$ holds for some positive
constants $C_2, C_3$ and all $y\in I_0$, $z\in K$.
\item{\em (a7)}
$\; v(y,\nu)=0 \;$ for all $y\in I_0$.
\item{\em (a8) }
There exists a zero-measure set $N\subset I_1$ and a number
$$\nu_1>d_1:=\sup_{x\in I_1}\, [x-u^{-1}(x)-\nu]>0$$ such that $$
\varrho(x)v(u^{-1}(x),\xi)> 0 $$ for all $x\in I_1\setminus N$ and
$\xi\in (\nu,\nu+\nu_1)$.
\end{description}
\begin{lemma}
\label{bounds} Assume (a1)--(a7). Then there exists a number
$C_{4}$ such that
\begin{equation} \label{Gbound}
|\GG(y,\xi\pm i\eta)|\leq C_{4}
\end{equation}
holds for all $y\in I_0$, $\xi\in\R$, and $0\not=\eta\in\R$.
\end{lemma}
{\sl Proof:} Recall the definition
$$ \GG(y,\xi\pm i\eta)
=\int_\nu^\infty\frac{v(y,z)}{y+z-\xi\mp i\eta}\,dz =
\int_\nu^\infty\frac{y+z-\xi\pm i\eta}{(y+z-\xi)^2+\eta^2}
v(y,z)\,dz \,. $$
Using the first part of {\em (a6)}, we get
$$ |\Im\GG(y,\xi\pm i\eta)|=
\int_\nu^\infty\frac{|\eta|v(y,z)}{(y+z-\xi)^2+\eta^2}\, dz \leq
C_2\int_{-\infty}^\infty\frac{|\eta|}{z^2+\eta^2}\, dz=\pi C_2 \,.
$$
We fix $\alpha>0$ and distinguish several cases. \\ [2mm]
{\em (i)} $\;\xi-y\geq\nu+\alpha\,$. Then
$$ \Re\GG(y,\xi\pm
i\eta)=\left(\int_{\nu+y-\xi}^{-\alpha}+\int_{-\alpha}^\alpha
+\int_\alpha^\infty\right)\frac{zv(y,z-y+\xi)}{z^2+\eta^2}\,dz $$
where by {\em (a2)} we have
$$ \left|\left(\int_{\nu+y-\xi}^{-\alpha}+
\int_\alpha^\infty\right)\frac{zv(y,z-y+\xi)}{z^2+\eta^2}\,dz\right|
\leq \frac{1}{\alpha}\int_\nu^\infty v(y,z)\,dz \leq
\frac{C}{\alpha} \,. $$
Using the mean value theorem,
\begin{eqnarray*}
J_2:&=&\int_{-\alpha}^{\alpha}\frac{zv(y,z-y+\xi)}{z^2+\eta^2}\,dz\\
&=&\int_{-\alpha}^\alpha\frac{z}{z^2+\eta^2}\left[v(y,\xi-y) +z
\partial_2 v(y,\vartheta(y,\xi,z))\right]\,dz
\end{eqnarray*}
with $\vartheta(y,\xi,z)$ between $\xi-y$ and $\xi-y+z$. The
integral of the first term is zero due to the antisymmetry in $z$
while the second term can be estimated by {\em (a6)} giving
$|J_2|\leq 2C_3\alpha$ and
$$ |\Re\GG(y,\xi\pm i\eta)|\leq C\alpha^{-1}+2C_3\alpha \,. $$
\\ [2mm]
{\em (ii)} $\;\nu\leq\xi-y<\nu+\alpha\,$. Then
$$ \Re\GG(y,\xi\pm i\eta)= \left(\int_{\nu+y-\xi}^{\xi-\nu-y}
+\int_{\xi-\nu-y}^\alpha +\int_\alpha^\infty\right)
\frac{zv(y,z-y+\xi)}{z^2+\eta^2}\,dz \,, $$
where
$$
\left|\int_{\nu+y-\xi}^{\xi-\nu-y}\frac{z}{z^2+\eta^2}v(y,z-y+\xi)\,dz\right|
\leq 2C_3(\xi-\nu-y)\leq 2 C_3\alpha $$
follows by the same procedure as for the integral $J_2$ in case
(i) and
$$
\left|\int_{\alpha}^{\infty}\frac{z}{z^2+\eta^2}v(y,z-y+\xi)\,dz\right|
\leq C\alpha^{-1} $$
due to {\em (a2)}. In the remaining integral,
$$ |v(y,z-y+\xi)|\leq C_3(z-y+\xi-\nu) $$
by {\em (a6)} and {\em (a7)}. Denoting for a while $A=\xi-\nu-y$,
we have now
\begin{eqnarray*}
\lefteqn{ \left|
\int_{\xi-\nu-y}^\alpha\frac{z}{z^2+\eta^2}v(y,z-y+\xi)\,dz\right|
\leq C_3\int_A^\alpha\frac{z^2+Az}{z^2+\eta^2}\,dz } \\ && =
C_3\left[\alpha-A+\frac{A}{2}{\ln}
\frac{\alpha^2+\eta^2}{A^2+\eta^2} -
|\eta|\left({\arctan}\frac{\alpha}{|\eta|}
-{\arctan}\frac{A}{|\eta|}\right)\right]\\ && \leq
C_3\left[\alpha+\frac{A}{2}{\ln}\frac{\alpha^2+\eta^2}{A^2+\eta^2}\right]
\end{eqnarray*}
taking into account that $0\leq A\leq \alpha$ in the last
inequality. Let us estimate the maximum of function
$$ f(A)=\frac{A}{2}\: {\ln}\frac{\alpha^2+\eta^2}{A^2+\eta^2} $$
in the mentioned interval of $A$. Clearly $f(0)=f(\alpha)=0$ and
$f(A)>0$ for $0<A<\alpha$. Hence $f$ has a maximum at some point
$A_0\in(0,\alpha)$ satisfying
$$ f'(A_0)=\frac{1}{2}\: {\ln}\frac{\alpha^2+\eta^2}{A_0^2+\eta^2}
- \frac{A_0^2}{A_0^2+\eta^2}=0 \,. $$
From the last equation,
$$ f(A_0)=\frac{A_0^3}{A_0^2+\eta^2} \leq A_0 \leq \alpha\,. $$
As a result,
$$ \left| \int_{\xi-\nu-y}^\alpha\frac{z}{z^2+\eta^2}\:
v(y,z-y+\xi)\,dz \right| \leq 2C_3\alpha $$
and
$$ |\Re\GG(y,\xi\pm i\eta)|\leq 4C_3\alpha + C\alpha^{-1} $$
\\ [2mm]
{\em (iii)} $\;\nu-\alpha \leq \xi-y < \nu\,$. Then
$$ |\Re\GG(y,\xi\pm i\eta)|= \left|\left(\int_{\nu+y-\xi}^\alpha +
\int_\alpha^\infty\right) {z\over z^2+\eta^2}\, v(y, z-y+\xi)\, dz
\right| \,. $$
Here the second integral is bounded by $C\alpha^{-1}$ and the first
one we estimate similarly as in the case (ii). Denoting here
$B=\nu+y-\xi\in(0,\alpha]$, we obtain
$$ \left|\int_{\nu+y-\xi}^\alpha\frac{z}{z^2+\eta^2}v(y,z-y+\xi)\,
dz\right| \leq C_3\int_B^\alpha\frac{z(z-B)}{z^2+\eta^2}\, dz \leq
C_3\alpha $$
and
$$ |\Re\GG(y,\xi\pm i\eta)| \leq C_3\alpha + C\alpha^{-1} \,. $$
\\ [2mm]
{\em (iv)} $\;\xi-y<\nu-\alpha\,$. Then
$$ |\Re\GG(y,\xi\pm i\eta)|= \left|\int_{\nu+y-\xi}^\infty
\frac{z}{z^2+\eta^2}v(y,z-y+\xi)\, dz\right| \leq C\alpha^{-1} \,.
$$
Summing up the discussion, we have found that in all the cases the
inequality
$$ |\Re\GG(y,\xi\pm i\eta)| \leq 4C_3\alpha + C\alpha^{-1} $$
holds. Minimizing the right-hand side with respect to $\alpha>0$,
we get
$$ |\Re\GG(y,\xi\pm i\eta)| \leq 4\sqrt{CC_3} $$
and
\begin{equation}
\label{eq:estimate1} |\GG(y,\xi\pm i\eta)| \leq \sqrt{16CC_3 +
\pi^2C_2^2} \,,
\end{equation}
what we set out to prove. \quad \QED \vspace{2mm}
\begin{theorem}
\label{uW} Assume (a1)--(a8).
Then there exists $\delta_2>0$ such
that for all $0<|\kappa|<\delta_2$ and $t\in\R$
\begin{equation} \label{timeevol}
\label{u1} \psi(t,x)= \UU(t,x)\psi_0(x)
\end{equation}
holds for almost every $\,x\in I_1$, where
\begin{eqnarray}
\label{u2} \UU(t,x)&\!=\!& \int_{\nu+u^{-1}(x)}^\infty W(x,\xi)
e^{-i\xi t}\, d\xi \,, \\ \nonumber W(x,\xi)&\!=\!&
\frac{\kappa^2\varrho(x)v(u^{-1}(x),\xi-u^{-1}(x))} {\left[x\!-\!
\xi \!-\!\kappa^2\varrho(x)I(u^{-1}(x),\xi)\right]^2 +
\pi^2\kappa^4\varrho(x)^2 v(u^{-1}(x),\xi\!-\!u^{-1}(x))^2}\,. \\
&& \label{u3}
\end{eqnarray}
\end{theorem}
{\sl Proof:} Let $\delta$, $\Delta$ and $\zeta(x,\kappa)
=\zeta_1(x,\kappa)-i\zeta_2(x,\kappa)$ be as in
Theorem~\ref{singsimple}. We first verify that
$\zeta_2(x,\kappa)>0$ for $x\in I_1\setminus N$ and
$$0<|\kappa|<\delta_2':=\min\left(\delta,\sqrt{\frac{\nu_1-d_1}{C_1
C_4}},\sqrt{\frac{d}{C_1 C_4}}\right)\,,$$
where
$$ d:=\min_{x\in I_1}\, [x-\nu-u^{-1}(x)] > 0 $$
by assumption {\em (a5)}. It is sufficient to show that
$\zeta(x,\kappa)$ is not real as we know that
$\zeta_2(x,\kappa)\geq 0$. By assumption {\em (a2)} and Lemma
\ref{bounds},
$$ |D_+(x,\kappa,\xi)|\geq |\xi-x|-\kappa^2 C_1 C_4$$
for real $\xi$ and there is no solution in $(-\infty,x-\kappa^2
C_1 C_4)\cup (x+\kappa^2 C_1 C_4,\infty)$. If
$\zeta(x,\kappa)=\xi$ then the imaginary part of the
Eq.~(\ref{resing}) reads
$$ \kappa^2\pi\varrho(x) v(u^{-1}(x),\xi-u^{-1}(x))=0\quad .$$
Thus there are no real solutions in
$(\nu+u^{-1}(x),\nu+\nu_1+u^{-1}(x))\supset
(x-d,\nu+\nu_1+u^{-1}(x))$ by assumption {\em (a8)}. For the
considered values of $\kappa$ the intervals without real solutions
$\xi$ cover the whole real axis.

To any natural
number $n$ there exists an open set $N_n\subset \R$ of Lebesgue
measure smaller then $\frac{1}{n}$ such that $N\subset
N_{n+1}\subset N_n$. Let us denote $I_n'=I_1\setminus N_n$. Let
$\varphi$ be an arbitrary vector from ${\mathcal{H}}_0$ and
$\varphi_n=\varphi\chi_{I_n'}$. The scalar product
\begin{equation}
\label{spt}
(\varphi_n,\psi(t,\cdot))=\lim_{\eta\to 0^+}\frac{1}{\pi}
\int_{I_1\times\R}
\overline{\varphi_n(x)}\left(\Im r(x,\xi+i\eta)\right)e^{-i\xi t}
\psi_0(x) w_1(x)\,dx\,d\xi
\end{equation}
with $r(x,\xi+i\eta)$ given by (\ref{rresolvform}). Fubini
theorem can be used here as
\begin{eqnarray*}
|\GG(u^{-1}(x),\xi+i\eta)|\leq\frac{C}{\eta}\,, &&
|r(x,\xi+i\eta)|\leq\frac{1}{\eta}\,, \\ |\Im r(x,\xi+i\eta)|
&\!=\!& {\cal O}(\xi^{-2})\quad{\mathrm{as}}\quad \xi\to\pm\infty
\end{eqnarray*}
 for
$\eta>0$ using {\em (a2)}, (\ref{fG}) and (\ref{rresolvform})
only.
Let us next choose $0<\Delta_1<\Delta$, $0<\eta_1<\Delta$, denote
\begin{equation}
\delta_2=\min\left(\delta_2',
\sqrt{\frac{\Delta_1}{2C_1C_4}}\right) \,,
\end{equation}
and consider further only $0<|\kappa|<\delta_2$, $0<\eta\leq\eta_1$.
Now
$$ |\kappa^2\varrho(x)\GG(u^{-1}(x),\xi+i\eta)|\leq
\frac{1}{2}\Delta_1 $$
by Lemma \ref{bounds}. We divide the integration range
$I_1\times\R$ of (\ref{spt}) into the parts where $|\xi-x|\geq
\Delta_1$ and $|\xi-x|\leq\Delta_1$, respectively, and construct
the integrable majorant allowing us to use the dominated
convergence in (\ref{spt}).

For $|\xi-x|\geq\Delta_1$, clearly
$$ |\Im r(x,\xi+i\eta)| \leq \frac{\eta_1+\frac{1}{2}\Delta_1}
{\left(|\xi-x|-\frac{1}{2}\Delta_1\right)^2} \leq
\frac{4\eta_1+2\Delta_1}{\Delta_1^2} \,. $$
Let us define function $g:\R\to\R$ as (recall that
$I_1=\big\lbrack\xi_1^{(-)}, \xi_1^{(+)} \big\rbrack$)
\begin{equation}
\label{gdef}
g(\xi)=\left\{
\begin{array}{lcr}
(\xi-\xi_1^{(-)} +\frac{\Delta_1}{2})^{-2}
(\eta_1+\frac{\Delta_1}{2})
& \dots &\xi
<\xi_1^{(-)}-\Delta_1\\ \frac{4\eta_1+2\Delta_1}{\Delta_1^2}&
\dots &\xi_1^{(-)}-\Delta_1\leq\xi\leq \xi_1^{(+)}
+\Delta_1\\ (\xi-\xi_1^{(+)} -\frac{\Delta_1}{2})^{-2}
(\eta_1+\frac{\Delta_1}{2})&
\dots &\xi> \xi_1^{(+)}+\Delta_1
\end{array}
\right.
\end{equation}
Then $(x,\xi) \mapsto g(\xi)|\varphi(x)|\, |\psi_0(x)|\, w_1(x)$
is the sought majorant.

For $|\xi-x|\leq \Delta_1$ we can consider only
$x\in I_n'$ as $\varphi_n(x)=0$
elsewhere. By Theorem \ref{singsimple},
$$ m_{\kappa,n}:=\min |D_{+}(x,\kappa,\xi+i\eta)| >0 $$
where $D_+$ is defined in (\ref{implf}) and the minimum is taken
over the considered set of variables $x\in I_n'$,
$\xi\in [x-\Delta_1,x+\Delta_1]$, $\eta\in
[0,\eta_1]$ and a fixed value of $0<|\kappa|<\delta_2$ (notice
that $\xi+i\eta\in\Omega_x$ due to our choice of $\Delta_1$,
$\eta_1$ and the inclusion in (\ref{resingint})). The
majorant can be now chosen as
$$ m_{\kappa,n}^{-1}|\varphi(x)|\, |\psi_0(x)|\, w_1(x) \,. $$

Interchanging the limit with the integral in (\ref{spt}), using
Lemma \ref{uplimlemma} and realizing that the integrand limit
vanishes for $\xi<\nu+u^{-1}(x)$, we obtain
\begin{equation}
(\varphi_n,\psi(t,\cdot))_{I_n'}=(\varphi_n,\UU(t,\cdot)\psi_0)_{I_n'}
\end{equation}
with $\UU$ given by (\ref{u2})-(\ref{u3}) and scalar products
in the space $L^2(I_n',w_1(x)\,dx)$.
Here $\UU(t,\cdot)\psi_0\in L^2(I_n',w_1(x)\,dx)$ as $\UU$ is bounded
in $\R\times I_n'$ which can be seen using the majorant
constructed above.
As $\varphi_n\in L^2(I_n',w_1(x)\,dx)$ may be arbitrary,
Eq. (\ref{u1}) follows for a.e. $x\in I_n'$. Now we see (\ref{u1})
for a.e. $x\in I_1$ in the limit $n\to\infty$.
\quad \QED \vspace{2mm}

\setcounter{equation}{0}
\renewcommand{\thesection}{\Roman{section}}
\section{Exponential decay at intermediate times}
\renewcommand{\thesection}{\arabic{section}}

Recall that decays of unstable quantum systems are nonexponential
at very short and very long times, however, they are usually
exponential in a very good approximation over a wide range of
intermediate times. Our aim here is to show that the present
models exhibits a similar behaviour in the sense that the function
$\UU(\cdot,x)$ appearing in the restricted time evolution operator
(\ref{timeevol}) can be approximated by an exponential for a.e.
fixed $x\in I_1$.

The way to prove that is inspired by \cite{DE}. We employ the fact
that the continued resolvent is for any fixed $x$ a meromorphic
function and show that for a sufficiently weak coupling the time
evolution is dominated by the contribution from the residue term
in (\ref{u2}).

In addition to the hypotheses made above, let us assume that
there exist a constant $C_5$ such that
\begin{description}
\item{\em (a9)}
$\;\left|\frac{\partial^2 v(y,z)}{\partial z^2}\right| \leq
C_5\;$ holds for all $y\in I_0$ and $z\in K$.
\end{description}

\begin{lemma}
\label{Gderestimate} For any $\alpha>-\nu$, $x\in I_1$, and
$\xi>u^{-1}(x)+\nu$ the following estimates hold:
\begin{eqnarray}
\nonumber \lefteqn{
\left|\Re\frac{\partial\GG_\Omega(u^{-1}(x),\xi)}{\partial\xi}\right|
\leq \left(\frac{C_2}{\alpha+\nu} +4C_3\right) }  \\ && +
C_3\left|\ln\frac{\xi-u^{-1}(x)-\nu}{\alpha+\nu}\right| +
C_5(\xi-u^{-1}(x)-\nu)\,, \label{GGRelog}
\end{eqnarray}
\begin{equation}
\label{GGImconst}
\left|\Im\frac{\partial\GG_\Omega(u^{-1}(x),\xi)}{\partial\xi}\right|
\leq \pi C_3\,. \phantom{AAAAAAAAAAAAAA}
\end{equation}
\end{lemma}
{\sl Proof:} Let us estimate
\begin{equation}
\frac{\partial\GG(u^{-1}(x),\zeta)}{\partial\zeta}=
\int_\nu^\infty\frac{v(u^{-1}(x),z)}{(u^{-1}(x)+z-\zeta)^2}\, dz
\end{equation}
for $\zeta=\xi+i\eta$, $\eta>0$, and get the result on the real
axis by taking the limit $\eta\to 0^{+}$ using
Lemma~\ref{analcont}. We rewrite the derivative as
\begin{equation}
\label{Gder1}
\frac{\partial\GG_\Omega(u^{-1}(x),\zeta)}{\partial\zeta} =
\int_{\nu+u^{-1}(x)-\xi}^\infty \frac{z^2-\eta^2+2i\eta
z}{(z^2+\eta^2)^2}\: v(u^{-1}(x),z+\xi-u^{-1}(x))\, dz
\end{equation}
and denote for a moment
\begin{equation}
\beta=\xi-u^{-1}(x)-\nu\,, \quad \gamma=\xi-u^{-1}(x)+\alpha \,;
\end{equation}
by assumption we have $0<\beta<\gamma$.

In the expression for the imaginary part of (\ref{Gder1}) we
separate the integrals over $(-\beta,\beta)$ and
$(\beta,\infty)$. In the second integral the limit $\eta\to 0$
gives zero as can be seen easily by the dominated convergence.
In the integral over $(-\beta,\beta)$, we insert the Taylor
expansion
\begin{eqnarray}
\lefteqn{ v(u^{-1}(x),z+\xi-u^{-1}(x)) =
v(u^{-1}(x),\xi-u^{-1}(x))\; + } \nonumber
\\ &&
\label{Taylorv} \frac{\partial
v(u^{-1}(x),\xi-u^{-1}(x))}{\partial z}\, z + \frac{1}{2}
\frac{\partial^2 v(u^{-1}(x),\xi-u^{-1}(x)+\theta)}{\partial
z^2}\, z^2 \phantom{AAA}
\end{eqnarray}
where $\theta$ in the error term lies between $0$ and $z$. The
contribution of the $z^0$ term to the integral vanishes because it
gives rise to an odd function. The contribution of the second term
is bounded by $\pi C_3$ in the limit $\eta\to 0$ as it follows
from assumption {\em (a6)} and an explicit calculation. The $z^2$
term again does not contribute in view of assumption {\em (a9)}
and an explicit calculation. In this way, inequality
(\ref{GGImconst}) is proved.

As for the real part of Eq. (\ref{Gder1}), we proceed similarly.
Inserting the expansion (\ref{Taylorv}) into the integral over
$(-\beta,\beta)$ we obtain from the $z^0$ term
$$ \left|-\frac{2}{\beta}\: v(u^{-1}(x),\xi-u^{-1}(x)) \right|
\leq 2 C_3\,, $$
where the assumptions {\em (a6)} and {\em (a7)} were used in the
last inequality. The term with $z$ does not contribute and the
term with $z^2$ is estimated by $C_5\beta$ in the limit $\eta\to
0$. The integral over $(\beta,\gamma)$ in (\ref{Gder1}) can be
handled by means of {\em (a6)} and {\em (a7)},
\begin{eqnarray*}
\lefteqn{ \left|\int_\beta^\gamma
\frac{z^2-\eta^2}{(z^2+\eta^2)^2}\:
v(u^{-1}(x),z+\xi-u^{-1}(x))\, dz \right| } 
\\ && = \left| \int_\beta^\gamma
\frac{z^2-\eta^2}{(z^2+\eta^2)^2}\: \frac{\partial
v(u^{-1}(x),\theta_1)}{\partial z} (z+\xi-u^{-1}(x)-\nu)\, dz
\right| 
\\ && \leq C_3 \int_\beta^\gamma \frac{z+\xi-u^{-1}(x)-\nu}{z^2}\, dz
= C_3 \left[\ln\left(1+\frac{\alpha+\nu}{\beta}\right) +
\frac{\alpha + \nu}{\gamma}\right] 
\\ && \leq C_3 \left[ 2 + \left|\, \ln\frac{\xi-u^{-1}(x)-\nu}{\alpha
+ \nu}\right|\right]\,,
\end{eqnarray*}
where we have employed $\nu<\theta_1<z+\xi-u^{-1}(x)$ and the
inequality $\ln (1+x) \leq 1 + |\ln x|$. Finally, we have
$$ \left| \int_\gamma^\infty \frac{z^2-\eta^2}{(z^2+\eta^2)^2}\:
v(u^{-1}(x),z+\xi-u^{-1}(x))\, dz \right| \leq C_2
\int_\gamma^\infty \frac{dz}{z^2} \leq \frac{C_2}{\alpha + \nu}
\,; $$
putting all these estimates together, we arrive at
(\ref{GGRelog}). \quad \QED \vspace{2mm}
\begin{lemma}
\label{Wabscont}
There is $\delta_3>0$ such that for all
$0<|\kappa|<\delta_3$ and almost every $x\in I_1$, the function
$W(x,\cdot)$ defined by formula (\ref{u3}) for
$\xi>\nu+u^{-1}(x)$ and extended by zero to the rest of the real
axis, $W(x,\xi)=0$ for $\xi\leq\nu+u^{-1}(x)$, is absolutely
continuous in any compact subinterval of $\,\R$.
\end{lemma}
{\sl Proof:} From the proof of Theorem \ref{uW} we know that
$$ W(x,\xi)=\frac{1}{\pi}\lim_{\eta\to 0^+}\Im r(x,\xi+i\eta)\,, $$
and therefore
\begin{eqnarray}
\label{WO} W(x,\xi) &\!=\!& \frac{1}{\pi}\Im r_\Omega(x,\xi) \,, \\
\label{rO} r_\Omega(x,\xi) &\! := \!&
\frac{1}{x-\xi-\kappa^2\varrho(x)\GG_\Omega(u^{-1}(x),\xi)}= [
D_+(x,\kappa,\xi)]^{-1}
\end{eqnarray}
for $\xi>\nu+u^{-1}(x)$. Let $\delta$ and $\Delta$ be the numbers
from Theorem~\ref{singsimple}. For $\kappa^2<\Delta
C_1^{-1}C_4^{-1}$, $D_+$ has no zeros if $|\xi-x|\geq\Delta$ (see
Lemmas~\ref{analcont} and \ref{bounds} and assumption {\em (a2)}).
On the other hand, for $|\xi-x|<\Delta$ and $0<|\kappa|<\delta_2$
real zeros can exist for at most zero-measure set of $x$ which we
neglect (see the proof of Theorem~\ref{uW}). Apart of it
$W(x,\cdot)$ has a continuous derivative in
$(\nu+u^{-1}(x),\infty)$ and therefore it is absolutely continuous
in any compact subinterval. Let us denote
$$ d:=\min_{x\in I_1}\, [x-\nu-u^{-1}(x)] > 0\,, $$
where the positivity follows from assumption {\em (a5)}. Then
$|D_+(x,\kappa,\xi)|>\frac{d}{3}$ for
$\nu+u^{-1}(x)<\xi<\nu+u^{-1}(x)+\frac{d}{3}$,
$\kappa^2<\frac{d}{3 C_1 C_4}$, and $\frac{\partial
W(x,\xi)}{\partial \xi}$ is bounded by an expression similar to
the r.h.s. of (\ref{GGRelog}) in the considered interval of $\xi$.
Due to the integrability of $|\ln(\xi-\nu-u^{-1}(x))|$ and the
estimate
$$ |W(x,\xi)|\leq \frac{3\pi C_3}{C_4 d}\: (\xi-\nu-u^{-1}(x)) $$
(see assumptions {\em (a6)--(a7)} and (\ref{GGOdef}))
$\,W(x,\cdot)$ is absolutely continuous in $[\nu+u^{-1}(x),
\nu+u^{-1}(x)+\frac{d}{3}]$. Consequently, it is absolutely
continuous in any compact subinterval of $\R$. Choosing
$$ \delta_3=\min\, \left(\delta,\delta_2, \sqrt{\frac{\Delta}{C_1 C_4}},
\sqrt{\frac{d}{3 C_1 C_4}} \right) \,. $$
we get the desired result. \quad \QED \vspace{2mm}
\begin{lemma}
\label{M1M2} There exists $\delta_4>0$ such that
\begin{eqnarray}
\label{M1} M_1 &:=& \max_{x\in I_1, |\kappa|\leq \delta_4}
\left|\varrho(x)\GG_\Omega(u^{-1}(x),\zeta(x,\kappa))\right| < \infty
\,,
\\
\label{M2} M_2 &:=& \max_{x\in I_1, |\kappa|\leq \delta_4} \left|
\varrho(x)
\frac{\partial\GG_\Omega(u^{-1}(x),\zeta(x,\kappa))}{\partial\zeta}\right|
<\infty\,,
\end{eqnarray}
and for all $|\kappa|<\delta_5:=
\min(\delta_4,(2M_2)^{-\frac{1}{2}})$, $x\in I_1$, we have
\begin{eqnarray*}
|\zeta_1(x,\kappa)-x|\leq 2 M_1\kappa^2 \,, && \qquad
0\leq\zeta_2(x,\kappa)\leq 2 M_1\kappa^2 \,,\\
|\zeta(x,\kappa)-x|\leq 2 M_1\kappa^2 \,, && \qquad
\left|\frac{\partial\zeta(x,\kappa)}{\partial\kappa^2}\right| \leq
2 M_1\,,
\end{eqnarray*}
where $\zeta(x,\kappa)=\zeta_1(x,\kappa)-i\zeta_2(x,\kappa)$ is
the function from Theorem \ref{singsimple}.
\end{lemma}
{\sl Proof:} By Theorem \ref{singsimple} $\zeta$ is uniformly
continuous in $I_1\times [-\frac{\delta}{2},\frac{\delta}{2}]$.
Hence there is $0<\delta_4\leq\frac{\delta}{2}$ such that for
$|\kappa|\leq\delta_4$ and all $x\in I_1$ we have
$|\zeta(x,\kappa)-x|<d$. Then $\zeta_1(x,\kappa)>\nu+u^{-1}(x)$
and the functions in the r.h.s of Eqs. (\ref{M1}), (\ref{M2}) are
continuous. Consequently, $M_1$, $M_2$ are finite. For
$|\kappa|<\delta_5$ we now have
$$ \nonumber \left|\frac{\partial\zeta(x,\kappa)}{\partial
\kappa^2}\right| =
\left|\frac{\varrho(x)\GG_\Omega(u^{-1}(x),\zeta(x,\kappa))}
{1+\kappa^2\varrho(x)\frac{\partial
\GG_\Omega(u^{-1}(x),\zeta(x,\kappa))} {\partial \zeta}} \right|
\leq \frac{M_1}{1-\kappa^2 M_2} \leq 2 M_1 $$
and the sought estimates on $\zeta(x,\kappa)-x$ follow. \quad \QED
\vspace{2mm}
\begin{lemma}
\label{FM3M4} Let $\alpha$ be a number such that
$$ 0<\alpha<d:=\inf_{x\in I_1}(x-u^{-1}(x)-\nu) \,, \quad \alpha
<{\mathrm{dist}\,}(I_1,\C\setminus\Omega)\,, $$
and let us denote
\begin{eqnarray*}
N_{\alpha,x}&:=&\{\vartheta\in\C\,|\; |\vartheta-x|\leq\alpha\}
\,,
\\ N_\alpha&:=&\{(x,\vartheta)\in I_1\times\C\,|\; x\in I_1,\;
|\vartheta-x|\leq\alpha\} \,.
\end{eqnarray*}
Then
\begin{description}
\item{(i)}
$\;N_\alpha\subset\{(x,\vartheta)\in I_1\times \C\,|\; \vartheta
\in\Omega\setminus (-\infty, u^{-1}(x)+\nu]\}$,
\item{(ii)}
if $\;x\in I_1$ and $\,\vartheta\in N_{\alpha,x}$ then
$(x,\vartheta)\in N_\alpha$,
\item{(iii)} $\;N_\alpha$ is closed in $\R\times\C$,
\item{(iv)} the numbers
\begin{eqnarray}
\label{M3} M_3(\alpha) &:=& \max_{(x,\vartheta)\in N_\alpha}
\left|\varrho(x)\GG_\Omega(u^{-1}(x),\vartheta)\right| < \infty \,,
\\
\label{M4} M_4(\alpha) &:=& \max_{(x,\vartheta)\in N_\alpha}
\left| \varrho(x)
\frac{\partial\GG_\Omega(u^{-1}(x),\vartheta)}{\partial\vartheta}\right|
<\infty
\end{eqnarray}
are finite,
\item{(v)} there exists an $\alpha'>\alpha$ such that for any $x\in I_1$,
$\vartheta\in N_{\alpha',x}$, and
$|\kappa|<\delta_6(\alpha):=\min\Big(\delta_5,
\sqrt{\frac{\alpha}{4M_1}}\Big)$ (see Lemma \ref{M1M2}) we have
\begin{eqnarray}
\nonumber \lefteqn{
\GG_\Omega(u^{-1}(x),\vartheta)=\GG_\Omega(u^{-1}(x),\zeta(x,\kappa))
} \\ && +\, \frac{\partial\GG_\Omega(u^{-1}(x),\zeta(x,\kappa))}
{\partial \zeta} (\vartheta - \zeta(x,\kappa)) + {\mathcal
F}(x,\vartheta) (\vartheta-\zeta(x,\kappa))^2 \phantom{AAAA}
\label{FFdef}
\end{eqnarray}
where ${\mathcal F}(x,\cdot)$ is a function holomorphic in the
interior of $N_{\alpha',x}$ and
\begin{eqnarray}
\label{M3est} \left|\varrho(x){\mathcal F}(x,\vartheta)\right| & \leq
& \frac{8 M_3(\alpha)}{\alpha^2}=:m_3(\alpha) \,, \\
\label{M33est} \left|\varrho(x)\frac{\partial {\mathcal
F}(x,\vartheta)}{\partial \vartheta}\right| & \leq & \frac{16
M_3(\alpha)}{\alpha^3}=:m_4(\alpha)
\end{eqnarray}
holds for $\vartheta$ in the interior of $N_{\alpha/2,x}$.
\end{description}
\end{lemma}
{\sl Proof:} The claims (i)--(iii) trivially follow from the
definitions, the claim (iv) follows from the assumption {\em (a4)}
and the claims (i), (iii). Under our assumptions there exists
$\alpha'>\alpha$ satisfying all the assumptions of the lemma. Then
for any $x\in I_1$, the function $\GG_\Omega(u^{-1}(x), \cdot)$ is
holomorphic in the interior of $N_{\alpha',x}$, the function
${\mathcal F}$ defined by Eq. (\ref{FFdef}) exists and ${\mathcal
F}(x,\cdot)$ is holomorphic in the interior of $N_{\alpha',x}$.
For $|\kappa|<\delta_6(\alpha)$ now $\zeta(x,\kappa)$ is in the
interior of $N_{\alpha,x}$ and for all $\vartheta$ in the interior
of $N_{\alpha,x}$ we have
\begin{eqnarray*}
\varrho(x){\mathcal F}(x,\vartheta)=\frac{1}{2\pi i}\int_{\partial
N_{\alpha,x}} \frac{\varrho(x)\GG_\Omega(u^{-1}(x),z)}
{(z-\zeta(x,\kappa))^2(z-\vartheta)}\: dz \,, \\
\varrho(x)\frac{\partial {\mathcal F}(x,\vartheta)}{\partial
\vartheta} = \frac{1}{2\pi i}\int_{\partial N_{\alpha,x}}
\frac{\varrho(x)\GG_\Omega(u^{-1}(x),z)}
{(z-\zeta(x,\kappa))^2(z-\vartheta)^2}\: dz\,.
\end{eqnarray*}
If $|\kappa|<\delta_6(\alpha)$ and $\vartheta\in
N_{\frac{\alpha}{2},x}$, then
\begin{eqnarray}
|z-\zeta(x,\kappa)|&\!\geq\!& |z-x|-|x-\zeta(x,\kappa)|\geq \alpha
- 2M_1\kappa^2 > \frac{\alpha}{2} \,,\\ |z-\vartheta| &\!\geq\!&
|z-x|-|x-\vartheta|\geq \frac{\alpha}{2}
\end{eqnarray}
by Lemma \ref{M1M2}, and the inequalities (\ref{M3est}),
(\ref{M33est}) follow immediately. \quad \QED \vspace{2mm}
\begin{theorem} \label{interexp}
Assume (a1)-(a9).
Then
there exist finite constants $\delta'>0$ and $C_6>0$ such that for
all $|\kappa|<\delta'$ and $t>0$ we have
\begin{equation}
\label{uexp} \left|\, \UU(t,x) -
A(x,\kappa)e^{-i\zeta_1(x,\kappa)t-\zeta_2(x,\kappa)t} \right|
\leq \frac{C_6\kappa^2}{t}
\end{equation}
for a.e. $x\in I_1$ where
$\zeta(x,\kappa)=\zeta_1(x,\kappa)-i\zeta_2(x,\kappa)$ is the
singularity location (with $\zeta_1$ real, $\zeta_2\geq 0$ --
cf.~Theorem~\ref{singsimple}) and
$$ A(x,\kappa):=\left\lbrack 1+\kappa^2\varrho(x)
\frac{\partial\GG_\Omega(u^{-1}(x),\zeta(x,\kappa))}{\partial
\zeta} \right\rbrack^{-1} \,. $$
\end{theorem}
{\sl Proof:} If $\kappa=0$ we have $\zeta(x,0)=x$ by
(\ref{resing}) and $\UU(t,x)=e^{-ixt}$ (see (\ref{H_0})) so the
theorem holds with any $C_6$. Let us further suppose that
$\kappa\not= 0$. By Theorem~\ref{singsimple} and assumption {\em
(a8)}, $\zeta_2(x,\kappa)>0$ for a.e. $x\in I_1$ if
$|\kappa|<\delta_2$. Let us exclude the remaining zero-measure set
of $x$'s from our considerations. Then the integral
\begin{equation}
\label{poleint} \int_{-\infty}^{\infty}e^{-i\xi t} V(x,\xi)
\, d\xi =\lim_{R\to\infty}\int_{-R}^R e^{-i\xi t} V(x,\xi)\,
d\xi\,,
\end{equation}
where
\begin{equation}
\label{Vdef} V(x,\xi)=\frac{1}{\pi}\,
\Im\,\frac{A(x,\kappa)}{\zeta(x,\kappa)-\xi}\,,
\end{equation}
exists in the generalized sense (\ref{poleint}). While the
Lebesgue integral does not exist due to the behavior at large
$|\xi|$, the existence of generalized integral is well known and
will be in fact seen from our calculations below. We shall
estimate the difference between $\UU(t,x)$ in Eq. (\ref{u2}) and
the integral (\ref{poleint}).

Let us recall from the proof of Theorem \ref{uW} that
\begin{equation}
\label{Wlim} W(x,\xi)=\lim_{\eta\to 0^+} \frac{1}{\pi}\,
\Im\, r(x,\xi+i\eta) = \frac{1}{\pi}\, \Im\,
\frac{1}{x-\xi-\kappa^2\varrho(x)\GG_\Omega(u^{-1}(x),\xi)}\,,
\end{equation}
where the last equality should be used for $\xi>\nu+u^{-1}(x)$
only. Combining this with (\ref{u3}), assumptions {\em (a2), (a6)}
and Lemma \ref{bounds} we arrive at the estimate
\begin{equation}
\label{Winfty} |W(x,\xi)|\leq \frac{\kappa^2 C_1
C_2}{(\xi-x-\kappa^2 C_1 C_4)^2}
\end{equation}
for $\xi>x+\kappa^2 C_1 C_4$. Due to Lemma~\ref{Wabscont} we can
integrate by parts for $|\kappa|<\delta_3$,
$$ \int_{-\infty}^\infty e^{-i\xi
t}[W(x,\xi)-V(x,\xi)]\, d\xi=
-\frac{i}{t}\int_{-\infty}^\infty e^{-i\xi t}
\frac{\partial}{\partial\xi}[W(x,\xi)-V(x,\xi)]\, d\xi
\,. $$
Let us choose an $\alpha>0$ satisfying the assumptions of Lemma
\ref{FM3M4} and consider only the values of the coupling constants
such that
\begin{equation}
\label{d1} 0<|\kappa|<\min \left(\delta_3,\delta_6(\alpha),
\frac{1}{2}\sqrt{\frac{\alpha}{C_1 C_4}} \right) \,.
\end{equation}
To calculate $\frac{\partial}{\partial\xi}W(x,\xi)$ let us denote
for a while
\begin{eqnarray*}
D_+=D_+(x,\kappa,\xi) &\!=\!&
x-\xi-\kappa^2\varrho(x)\GG_\Omega(u^{-1}(x),\xi) \,,\\ D_1=\Re D_+
&\!=\!& x-\xi-\kappa^2\varrho(x)\Re\GG_\Omega(u^{-1}(x),\xi) \,,\\
D_2=\Im D_+ &\!=\!& -\kappa^2\varrho(x) \Im\GG_\Omega(u^{-1}(x),\xi)
\,,
\\ D_1'= \frac{\partial D_1}{\partial\xi} &\!=\!&
-1-\kappa^2\varrho(x)
\Re\frac{\partial}{\partial\xi}\GG_\Omega(u^{-1}(x),\xi) \,,\\
D_2'=\frac{\partial D_2}{\partial\xi} &\!=\!& -\kappa^2\varrho(x)
\Im\frac{\partial}{\partial\xi}\GG_\Omega(u^{-1}(x),\xi) \,.
\end{eqnarray*}
Then
$$ \frac{\partial}{\partial\xi}W(x,\xi)= -\pi^{-1}|D_+|^{-4}
\left[(D_1^2-D_2^2)D_2'-2D_1D_1'D_2\right] \,. $$
If now $|\xi-x|\geq\frac{\alpha}{2}$ the assumption {\em (a2)}
together with Lemmas \ref{bounds} and \ref{Gderestimate} (where we
denote the constant as $C_3'$) imply
\begin{eqnarray*}
|D_+|&\geq& |\xi-x|-\kappa^2 C_1 C_4>\frac{\alpha}{4}>0 \,,\\
|D_+|&\geq& \frac{1}{2}|\xi-x| \,,\\ |D_1|&\leq& |\xi-x|+\kappa^2
C_1 C_4<2|\xi-x| \,,\\ |D_2|&\leq&\kappa^2 C_1
C_4<\frac{\alpha}{4} \,,\\ \nonumber |D_1'|&\leq& C_3'+\kappa^2
C_1 C_3 |\ln(\xi-u^{-1}(x)-\nu)| + \kappa^2 C_1 C_5
(\xi-u^{-1}(x)-\nu) \,,\\ |D_2'|&\leq& \kappa^2 \pi C_1 C_3 \,.
\end{eqnarray*}
From here we get
$$ \int_{(\nu+u^{-1}(x),x-\frac{\alpha}{2})\cup
(x+\frac{\alpha}{2},\infty)}
|\frac{\partial}{\partial\xi}W(x,\xi)|\, d\xi \leq C_7\kappa^2\,,
$$
where the explicit value of the constant $C_7$ can be expressed
from the above estimates if necessary. What is important is that
$C_7$ can be chosen independent of $\kappa$ in the considered
range.

Let us consider the term $V(x,\xi)$ now. We have the bounds
$$ \frac{1}{1+\kappa^2 M_2}\leq |A(x,\kappa)| \leq
\frac{1}{1-\kappa^2 M_2} $$
by Lemma \ref{M1M2}, so
\begin{equation}
\label{Aestim} \frac{2}{3}\leq |A(x,\kappa)|\leq 2
\end{equation}
holds for
\begin{equation}
\label{d2} |\kappa|<\min \left(\delta_4,(2M_2)^{-1/2} \right) \,.
\end{equation}
Denoting for a while $A_1=\Re A(x,\kappa)$, $A_2=\Im A(x,\kappa)$,
we have
$$ |A_1|\leq |A(x,\xi)|\leq 2\,, \quad |A_2|\leq \kappa^2
M_2 |A(x,\kappa)|^2\leq 4\kappa^2 M_2 $$
and
$$ \frac{\partial}{\partial\xi}V(x,\xi)=\frac{1}{\pi}\,
\frac{A_2[(\xi-\zeta_1(x,\kappa))^2-\zeta_2(x,\kappa)^2] -
2A_1\zeta_2(x,\kappa)(\xi-\zeta_1(x,\kappa))}
{[(\xi-\zeta_1(x,\kappa))^2 + \zeta_2(x,\kappa)^2]^2} \,. $$
If $|\xi-x|\geq \frac{\alpha}{2}$ and
\begin{equation}
\label{d3}
|\kappa|<\min\left(\delta_5,\sqrt{\frac{\alpha}{8M_1}}\right)\,,
\end{equation}
we have $|\xi-\zeta_1(x,\kappa)|\geq \frac{\alpha}{4}$ by Lemma
\ref{M1M2}, and therefore
$$ \int_{(-\infty,x-\frac{\alpha}{2})\cup
(x+\frac{\alpha}{2},\infty)} \left|
\frac{\partial}{\partial\xi}V(x,\xi) \right|\, d\xi \leq
C_8\kappa^2 $$
with a $\kappa$-independent finite constant $C_8$ which can be
given explicitly if necessary.

Let us now turn to $\xi\in (x-\frac{\alpha}{2},
x+\frac{\alpha}{2})$. Using the expansion (\ref{FFdef}),
\begin{eqnarray*}
W(x,\xi)-V(x,\xi) &\!=\!& \frac{1}{\pi}\: \Im \frac{\kappa^2
A(x,\kappa)^2\varrho(x){\mathcal F}(x,\xi)} {1+\kappa^2
A(x,\kappa)\varrho(x)(\xi-\zeta(x,\kappa)) {\mathcal F}(x,\xi)} \,,
\\ \frac{\partial}{\partial\xi}[W(x,\xi)-V(x,\xi)] &\!=\!&
\frac{1}{\pi}\: \kappa^2 \Im\Biggl\{ A(x,\kappa)^2 \Biggr. \\ &&
\Biggl. \times\,
\frac{\varrho(x)\frac{\partial{\mathcal F}(x,\xi)}{\partial\xi} -
\kappa^2 A(x,\kappa)\varrho(x)^2 {\mathcal F}(x,\xi)^2} {[1+\kappa^2
A(x,\kappa)\varrho(x)(\xi-\zeta(x,\kappa)) {\mathcal F}(x,\xi)]^2}
\Biggr\}\,.
\end{eqnarray*}
Using (\ref{Aestim}), (\ref{M3est}), (\ref{M33est}) together with
Lemma \ref{M1M2}, and assuming that
\begin{equation}
\label{d4} |\kappa|< \min \left(\delta_6(\alpha),
\frac{1}{2}\sqrt{\frac{\alpha}{M_1}}\,,
\frac{1}{2\sqrt{m_3(\alpha)\alpha}} \right)\,,
\end{equation}
we obtain
$$ \left|\frac{\partial}{\partial\xi}[W(x,\xi)-V(x,\xi)]
\right|\leq C_9\kappa^2\,, $$
where
$$ C_9:=\frac{16}{\pi}
\left(m_4(\alpha)+2\delta_6(\alpha)^2m_3(\alpha)^2
\right) \,. $$
Putting all the estimates together, we get
$$ \left|\int_{-\infty}^\infty e^{-i\xi t}\:
\frac{\partial}{\partial\xi}[W(x,\xi)-V(x,\xi)] \, d\xi \right|
\leq C_6\kappa^2\,, $$
where
$$ C_6=C_7+C_8+C_9\alpha $$
and
$$ |\kappa|<\delta'\,; $$
$\delta'$ being the minimum of $\delta$, $\delta_2$ and the r.h.s. in
(\ref{d1}),(\ref{d2})--(\ref{d4}). Evaluating the generalized integral
$$ \int_{-\infty}^{\infty} e^{-i\xi t}\, V(x,\xi)\, d\xi =
A(x,\kappa) e^{-i\zeta(x,\kappa)t} $$
by closing the integration contour in the lower half-plane for
$t>0$ the inequality (\ref{uexp}) is obtained. \quad \QED
\vspace{2mm}

The theorem is apparently useless for very short and very large
times when the error estimate $O(\kappa^2t^{-1})$ is much larger
then the amplitude value $\approx \exp(-\zeta_2(x,\kappa)t)$. On
the other hand, we get a nontrivial bound for the times when
\begin{equation}
\label{app1} \frac{C_6\kappa^2}{t}\ll e^{-\zeta_2(x,\kappa)t}
\end{equation}
where we take into account that $A(x,\kappa)\approx 1$. Let us
write
\begin{eqnarray}
\zeta_2(x,\kappa) &\!=\!& \kappa^2\eta_2(x,\kappa) \,, \nonumber
\\ \label{eta2} \eta_2(x,\kappa) &\!=\!&
\pi\varrho(x)v(u^{-1}(x),x-u^{-1}(x)) + O(\kappa^2)
\end{eqnarray}
for small coupling $\kappa$. In the subsequent formulas we do not
write the arguments of $\eta_2$, however, its $x$-dependence
should be kept in mind in general. The relation (\ref{app1}) is
valid for $T_1\ll t \ll T_2$ where $T_1$, $T_2$ are two solutions
of the equation
\begin{equation}
\label{T1T2} \kappa^2\eta_2 T_i  e^{-\kappa^2\eta_2 T_i} =
C_6\kappa^4\eta_2 \,, \quad i=1,2.
\end{equation}
If $\kappa^2 T_1$ is small we can approximate the equation by
replacing the exponential with one obtaining
$$ T_1\approx C_6\kappa^2 \,. $$
On the other hand, if $\kappa^2\eta_2T_2\gg 1$ we do not enlarge
the range $(T_1,T_2)$ by dropping the linear factor in
(\ref{T1T2}). Then we obtain
$$ T_2\approx -\frac{1}{\kappa^2\eta_2} \ln(C_6\kappa^4\eta_2)\,.
$$
The r.h.s. here is an decreasing function of $\eta_2$ in the
interval $(0,C_6^{-1}\kappa^{-4})$. By (\ref{eta2}) and
assumptions {\em (a2), (a6)} we have
$$ 0\leq\eta_2\leq \pi C_1 C_2 $$
in the $\kappa^0$ approximation. Restricting ourselves then to the
coupling constant values with
$$ |\kappa|\ll (\pi C_1 C_2 C_6)^{-1/4}\,, $$
we can safely use
$$ T_2\approx -\frac{1}{\pi C_1 C_2 \kappa^2} \ln(\pi C_1 C_2
C_6\kappa^4) \,. $$
Hence we see that the announced approximately exponential
behaviour of $\UU(\cdot,x)$ holds in the weak-coupling regime over
wide time range, roughly speaking from
$C_1^{-\frac{1}{4}}C_2^{-\frac{1}{4}}C_6^{\frac{3}{4}}\kappa$
to $\;C_1^{-1}C_2^{-1}\kappa^{-2}$.

\setcounter{equation}{0}
\renewcommand{\thesection}{\Roman{section}}
\section{Long time behavior}
\renewcommand{\thesection}{\arabic{section}}

The fact that the bound given by Theorem~\ref{interexp} becomes
useless at very large times is not coincidental, because the decay
rate is indeed slower there. To illustrate this claim, for
instance, let $x\in I_1$ be such that
by Lemma \ref{analcont}, Theorem \ref{singsimple}(b), Theorem
\ref{uW} and assumptions {\em (a6)--(a8)}, we have
\begin{eqnarray}
\mbox{$W(x , \xi)$ is finite and continuous w.r.t. $\xi \in
\left[\left. \nu + u^{-1}\left( x\right) , \infty\right)\right.$}
\label{eq:8-21-1}
\end{eqnarray}
for $0<|\kappa| < \delta_{2}$, where
$\delta_{2}$ is the number from Theorem \ref{uW}. This holds for
a.e. $x\in I_1$.

 By {\em (a6)} and (\ref{u3}), we get
$$|W(x,\xi)| \leq \frac{\kappa^{2}\varrho(x)C_{2}}{\left\lbrack x -
\xi - \kappa^{2}\varrho(x)I(u^{-1}(x),\xi)\right\rbrack^2} =:
T(\xi).$$
Since $\lim_{\xi\to\infty}I(u^{-1}(x),\xi) = 0$ by (\ref{pvconv}),
we get
\begin{eqnarray} T(\xi) \mathop{\sim}_{\xi\to\infty}
\frac{\kappa^{2}\varrho(x)C_{2}}{(x - \xi)^{2}}. \label{eq:8-21-2}
\end{eqnarray}
Thus, $g(\xi) := \chi_{\left[\left.\nu +
u^{-1}(x),\infty\right.\right)}(\xi)W(x,\xi)$ is in $L^{2}(\R)$,
but its support is not the whole $\R$, and
$${\cal U}(t,x) = \int_{\nu + u^{-1}(x)}^{\infty}g(\xi)e^{-i\xi
t}d\xi$$
by (\ref{u2}). Applying now \cite[Corollary C2]{Ar3}, we find that
for a.e. $x \in I_{1}$ and $|\kappa| < \delta_{2}$
\begin{eqnarray}
\mbox{${\cal U}(t,x)$ does not decay exponentially as $|t| \to
\infty$.} \label{eq:general} \end{eqnarray}
To learn more about the long-time asymptotic behavior of ${\cal
U}(t , x)$, we adopt the conditions {\em (a10)--(a13)} below, and
employ the results of \cite{Ar3} and \cite{Ar2} in the same way as
in \cite[Thm 3.2(ii)]{Hi}.

Given $\nu \ge 0$ and $\theta \in (0 , \pi / 2)$, we define ${\bf
D}_{\nu,\theta}$ by
\begin{eqnarray}
{\bf D}_{\nu,\theta} := \left\{ \zeta \in  \C\, |\, \Re \zeta >
\nu,\, - \theta < \arg\zeta < 0\right\}.
\end{eqnarray}
If $\nu < \nu'$, we have therefore
\begin{eqnarray}
{\bf D}_{\nu' ,\theta} \subset {\bf D}_{\nu,\theta}.
\end{eqnarray}
Let us denote
\begin{equation}
\Omega(v):=\cap_{y\in I_0}(\Omega-y)\, .
\end{equation}
Notice that $\Omega(v)\subset\cap_{y\in I_0}\Omega_{v,y}$ by {\em
(a3)}. We shall assume:
\begin{description}
\item{\em (a10)} There exists $\theta_{0}
\in (0 , \pi /4)$ such that $\overline{{\bf D}_{\nu,\theta_{0}}}
\subset \Omega(v)$.
\item{\em (a11)} $\;v(y,\xi)>0$ holds for each $y \in I_0$ and
$\xi>\nu$.
\item{\em (a12)} Given $y \in  I_0$,
there exists $C_{y} > 0$ and $q_{y} > 0$ such that
$$|v\left( y , \zeta\right)| < C_{y}|\zeta|^{- q_{y}}$$
holds for any $\zeta \in \Omega(v)$.
\end{description}
Notice that for $v$ which is continuous by {\em (a3)}, the
assumption {\em (a11)} implies, in particular, that for each $x
\in I_1$ and $\alpha, \beta \in \left( \nu , \infty\right)$ we
have
\begin{equation} \label{old11}
\mu_{x,\alpha,\beta} := \inf_{\alpha \leq \xi \leq \beta} v\left(
u^{-1}\left( x\right) , \xi\right) >0 \,.
\end{equation}
For fixed $x \in  I_1$ and $\zeta \in  \Omega\setminus
\left(\left. - \infty , u^{-1}\left( x\right) +
\nu\right]\right.$, we have defined $D_{+}(x,\kappa,\zeta)$  by
(\ref{eq:D+}). In a similar way, we define three other functions,
$D_{-}(x,\kappa,\zeta)$, $W(x , \zeta)$, and $g_{x}(\zeta)$ by
\begin{eqnarray}
\label{Dminus}
D_{-}(x,\kappa,\zeta) &:=& x - \zeta - \kappa^{2}\varrho(x)
\GG^{\Omega}(u^{-1}(x) , \zeta)\,, \label{eq:D-} \\ \nonumber &{}&
\qquad \\ W(x , \zeta) &:=& \frac{\kappa^{2}\varrho(x)v\left(
u^{-1}\left( x\right) , \zeta - u^{-1}\left( x
\right)\right)}{D_{+}(x,\kappa,\zeta)D_{-}(x,\kappa,\zeta)}\,,
\label{eq:W} \\ \nonumber &{}& \qquad  \\ \nonumber g_{x}(\zeta)
&:=& W\left( x , \zeta + u^{-1}\left( x\right)\right) \\ &=&
\frac{\kappa^{2}\varrho(x)v\left( u^{-1}\left( x\right) ,
\zeta\right)}{D_{+}\left( x, \kappa , \zeta + u^{-1}\left(
x\right)\right) D_{-}\left( x , \kappa , \zeta + u^{-1}\left(
x\right)\right)}\,; \label{eq:*g}
\end{eqnarray}
in the last case $\zeta\in \Omega(v) \setminus (-\infty,\nu]$. Then,
for a.e. $x \in I_{1}$ and $\kappa \in \R$ with $0<|\kappa| <
\delta_{2}$,
\begin{eqnarray} \mbox{$g_{x}$ can be regarded as
measurable with $g_x \in L^{1}\left(\left(\nu , \infty\right) ,
d\xi\right)$} \label{eq:8-21-3}
\end{eqnarray}
by (\ref{eq:8-21-1}) and (\ref{eq:8-21-2}), and we can write the
time evolution as follows,
\begin{equation}
\label{eq:*u} {\cal U}(t , x) = e^{- i u^{-1}(x)t}
\int_{\nu}^{\infty} g_{x}(\xi)e^{-i\xi t}d\xi\,, \end{equation}
by (\ref{u2}) and (\ref{u3}).

Next we need several lemmas. The first of them follows from {\em
(a3), (a10),} and Lemma \ref{analcont}:
\begin{lemma}
\label{lem:A-g-m} $g_x(\zeta)$ is meromorphic in ${\bf
D}_{\nu,\theta_{0}}$ for every $x \in I_{1}$ and $\kappa \in \R$.
\end{lemma}
\begin{lemma}
\label{lem:A-g-a}
For every $x \in  I_1$ with $\varrho(x)\ne 0$, $\xi \in \R$ with $\xi
> \nu$, and $\kappa \in  \R$ with $0<|\kappa| <
\delta_{2}$,
\begin{eqnarray}
\lim_{\varepsilon\to 0^{+}} g_{x}(\xi - i\varepsilon) =
g_{x}(\xi). \label{eq:A-g-a}
\end{eqnarray}
\end{lemma}
{\sl Proof:} Let $\xi' \equiv \xi + u^{-1}(x)$. By Lemma
\ref{analcont} we have
\begin{eqnarray}
\label{eq:limD+} D_{+}(x,\kappa,\xi') &:=& \lim_{\varepsilon
\to 0^{+}} D_{+}(x,\kappa,\xi' - i\varepsilon) \\ \nonumber &=&
x - \xi' - \kappa^{2}\varrho(x) \Bigl\{ I\left( u^{-1}\left( x\right)
, \xi'\right) + i\pi v\left( u^{-1}\left( x\right) ,
\xi\right)\Bigr\},
\\ \label{eq:limD-} D_{-}(x,\kappa,\xi') &:=& \lim_{\varepsilon
\to 0^+} D_{-}(x,\kappa,\xi' - i\varepsilon)
\\ \nonumber &=& x - \xi' - \kappa^{2}\varrho(x) \Bigl\{ I\left(
u^{-1}\left( x\right) , \xi'\right) - i\pi v\left( u^{-1}\left(
x\right) , \xi \right)\Bigr\}
\end{eqnarray}
for $\xi' > u^{-1}(x) + \nu$\, ($\xi > \nu$), which implies that
\begin{eqnarray}
\label{eq:D+D-} \lefteqn{ D_{+}(x,\kappa,\xi')D_{-}(x,\kappa,\xi')
} \\ && = \left[ x - \xi' - \kappa^{2}\varrho(x) I\left( u^{-1}\left(
x\right) , \xi'\right) \right]^{2} + \pi^2\kappa^{4}\varrho(x)^{2} v\left(
u^{-1}\left( x\right) , \xi\right)^{2}. \nonumber
\end{eqnarray}
Then $\lim_{\varepsilon\to 0^+} W(x , \xi' - i\varepsilon) =
W(x , \xi')$ follows from (\ref{u3}) giving (\ref{eq:A-g-a}).
\quad \QED \vspace{2mm}
\begin{lemma}
\label{lem:A-g-b}
For every $x \in  I_{1}$,
with $\varrho(x) \ne 0$,
all sufficiently small $\varepsilon
> 0$, every $\alpha, \beta \in ( \nu , \infty)$ with $\alpha < \beta$,
and every $\kappa
\in \R$ with $0<|\kappa| < \delta_{2}$, there exists a constant
$C_{x, \alpha, \beta} > 0$ independent of $\varepsilon$ such that
\begin{eqnarray}
\sup_{\alpha < \xi < \beta} |g_{x}(\xi - i\varepsilon)| \leq C_{x,
\alpha, \beta}. \label{eq:A-g-b}
\end{eqnarray}
\end{lemma}
{\sl Proof:} Set $S_{p,q} := \left\{ \zeta \in  \C\, |\, p \leq
\Re\zeta \leq  q,\, -\nu\tan\theta_{0}\leq\Im\zeta\leq 0
\right\}$. Fix $\varepsilon' \in \R$ with $0 < \varepsilon' < 1$
arbitrarily. $v\left( u^{-1}\left( x\right) , \cdot\right)$ is
uniformly continuous in $S_{\alpha,\beta}$ by {\em (a3)} and {\em
(a10)} since $S_{\alpha,\beta} \subset \overline{{\bf
D}_{\nu,\theta_{0}}}$. So there exists a constant $\varepsilon_{1}
\equiv \varepsilon_{1}(x , \varepsilon')
> 0$ such that
$$\biggl|v\left( u^{-1}\left( x\right) , \xi - i\varepsilon\right)
-
v\left( u^{-1}\left( x\right) , \xi\right)\biggr| \leq
\varepsilon' \biggl|v\left( u^{-1}\left( x\right) ,
\xi\right)\biggr|$$
for $\alpha \leq \xi \leq \beta$ and $0 < \varepsilon <
\varepsilon_{1}$ and we have
\begin{eqnarray}
\biggl| v\left( u^{-1}\left( x\right) , \xi -
i\varepsilon\right)\biggr| \leq (1 + \varepsilon') \biggl| v\left(
u^{-1}\left( x\right) , \xi\right)\biggr| \label{eq:(2)}
\end{eqnarray}
for $\alpha \leq \xi \leq \beta$ and $0 < \varepsilon <
\varepsilon_{1}$. Since $D_{\pm}(x,\kappa,\cdot)$ is holomorphic
in $\Omega\setminus \left(\left. - \infty , u^{-1}(x) +
\nu\right]\right.$ by Lemma \ref{analcont},
$D_{\pm}(x,\kappa,\cdot)$ is uniformly continuous in
$S_{\alpha,\beta}+u^{-1}(x)$. In view of (\ref{eq:limD+})
and (\ref{eq:limD-}) there exists $\varepsilon_{2} \equiv
\varepsilon_{2}(x,\varepsilon') > 0$ such that
$$| D_{\pm}(x,\kappa,\xi' - i\varepsilon) -
D_{\pm}(x,\kappa,\xi')| \leq \varepsilon'
|D_{\pm}(x,\kappa,\xi')|$$
for $\xi' \equiv \xi + u^{-1}(x)$ with $\alpha \leq \xi \leq
\beta$ and $0 < \varepsilon < \varepsilon_{2}$. Hence we have
\begin{eqnarray}
(1 - \varepsilon')\biggl|D_{\pm}\left( x,\kappa,\xi + u^{-1}\left(
x\right)\right)\biggr| \leq \biggl|D_{\pm}\left( x,\kappa,\xi +
u^{-1}\left( x\right) - i\varepsilon\right)\biggr| \label{eq:(3)}
\end{eqnarray}
if $\alpha \leq \xi \leq \beta$ and $0 < \varepsilon <
\varepsilon_{2}$. Using further (\ref{old11}), (\ref{eq:*g}),
(\ref{eq:D+D-}), (\ref{eq:(2)}), and (\ref{eq:(3)}), we get
\begin{eqnarray*}
|g_{x}(\xi - i\varepsilon)| &\leq& \frac{(1 +
\varepsilon')\kappa^{2}\varrho(x) |v\left( u^{-1}\left( x\right) ,
\xi\right)|}{(1 - \varepsilon')^{2} |D_{+}\left( x,\kappa,\xi +
u^{-1}\left( x\right)\right) D_{-}\left( x,\kappa,\xi +
u^{-1}\left( x\right)\right)|}
\\
&{}& \qquad \\
 &\leq& \frac{(1 + \varepsilon')\kappa^{2}\varrho(x)
|v\left( u^{-1}\left( x\right) , \xi\right)|} {(1 -
\varepsilon')^{2}\pi^{2} \kappa^{4}\varrho(x)^{2} |v\left(
u^{-1}\left( x\right) , \xi\right)|^{2}}
\\
&{}& \qquad \\ &\leq& \frac{(1 + \varepsilon')}{(1 -
\varepsilon')^{2}
\mu_{x,\alpha,\beta}\pi^{2}\kappa^{2}\varrho(x)}
\end{eqnarray*}
for $\alpha \leq \xi \leq \beta$ and $0 < \varepsilon <
\varepsilon_{0} \equiv \min\left\{
\varepsilon_{1},\varepsilon_{2}\right\}$, which implies the
desired result. \quad \QED \vspace{2mm}
\begin{lemma}
\label{lem:A-g-c}
For every $x \in  I_1$, all sufficiently large
$|\zeta|$ with $\zeta \in  {\bf D}_{\nu,\theta_{0}}$, and every
$\kappa \in \R$ satisfying $0<|\kappa| < \delta_{2}$,
$$|g_{x}\left( \zeta\right)| \leq \frac{C_{10}}{|\zeta|^{2 + q_{y}}}$$
with a constant $C_{10} > 0$ independent of $\zeta \in {\bf
D}_{\nu,\theta_{0}}$.
\end{lemma}
{\sl Proof:} In this proof, we set $y=u^{-1}(x)$, $\xi' \equiv \xi
+ u^{-1}(x)$, and let $\xi>\nu\geq 0$.
Since
$$D_{-}(x,\kappa,\xi' -i\varepsilon) = x - (\xi' -
i\varepsilon) - \kappa^{2}\varrho(x)\GG (y , \xi' - i\varepsilon)$$
for every $\varepsilon > 0$, we get
$$ |D_{-}\left(x,\kappa,\xi +u^{-1}\left( x\right)
-i\varepsilon\right)|^{2} \geq \left( \xi +
A_{\varepsilon,x}\left( \xi\right) \right)^{2}, $$
where
$$A_{\varepsilon,x}\left( \xi\right) := \kappa^{2}\varrho(x)\Re\GG (y
, \xi' - i\varepsilon) + u^{-1}(x) - x\,.$$
Set
$$B_{x} := \kappa^{2}\varrho(x)C_{4} + |u^{-1}(x)| + |x| > 0\,.$$
Then we get $|A_{\varepsilon,x}(\xi)| \leq  B_{x}$ by Lemma
\ref{bounds}. Since we now take $\xi > 0$, we get for every
$C_{-}$ with $0 < C_{-} < 1$,
\begin{eqnarray*}
\left( \xi + A_{\varepsilon,x}\left( \xi\right) \right)^{2} -
C_{-}^{2}\xi^{2} &\geq& \xi^{2} -2B_{x}\xi
- C_{-}^{2}\xi^{2} \\ &=& (1 - C_{-}^{2}) \left( \xi -
\frac{B_{x}}{1 - C_{-}^{2}}\right)^{2}
- \frac{B_{x}^2}{1-C_{-}^2}\,.
\end{eqnarray*}
Thus there exists $C_{-}$ with $0 < C_{-} < 1$ and $\xi_{-} \equiv
\xi_{-}(x) > 0$ independent of $\varepsilon > 0$ such that
\begin{eqnarray}
\biggl|D_{-}\left(x,\kappa,\xi + u^{-1}\left( x\right) -
i\varepsilon\right)\biggr|
> C_{-}\xi
\label{eq:7-27-1}
\end{eqnarray}
for every $\xi > \xi_{-}$. As for $D_{+}(x, \kappa , \xi' -
i\varepsilon)$, we have
$$D_{+}(x,\kappa,\xi' -i\varepsilon) = D_{-}(x,\kappa,\xi'
-i\varepsilon) - 2i\kappa^{2}\varrho(x)\pi v\left( y , \xi -
i\varepsilon\right)$$
for any $\varepsilon > 0$. Moreover, by {\em (a12)} we get
$$|v(y,\xi - i\varepsilon)| \leq \frac{C_{y}}{\left\{ \xi^{2} +
\varepsilon^{2}\right\}^{q_{y}/2}} \leq
\frac{C_{y}}{\xi^{q_{y}}}$$
for $\xi-i\varepsilon\in\Omega(v)$.
Thus there exists $\xi_{+}'> 0$ independent of $\varepsilon > 0$
such that if $\xi > \xi_{+}'$, then
$$|v(y , \xi - i\varepsilon)| <
\frac{C_{-}}{4\pi\kappa^{2}\varrho(x)}\,.$$
Together we get
\begin{eqnarray*}
\biggl|D_{+}\left(x,\kappa,\xi + u^{-1}\left( x\right) -
i\varepsilon\right)\biggr| &\geq& \biggl|D_{-}\left(x,\kappa,\xi +
u^{-1}\left( x\right) - i\varepsilon\right)\biggr| -
\frac{C_{-}}{2}
\\ &\geq& C_{-}\xi - \frac{C_{-}}{2} > 0
\end{eqnarray*}
for $\xi \geq \max\left\{\xi_{-},\xi_{+}',1\right\} =: \xi_{+}$ by
(\ref{eq:7-27-1}); notice that $\xi_{+}$ is independent of
$\varepsilon > 0$. On the other hand, we get
$$C_{-}\xi - \frac{C_{-}}{2} \geq \frac{C_{-}}{2}\,\xi$$
for $\xi > \xi_{+}$. Now we set $C_{+} := C_{-}/2$; then $0 <
C_{+} < 1$ and
\begin{eqnarray}
\biggl|D_{+}\left(x,\kappa,\xi +u^{-1}\left( x\right) -
i\varepsilon\right)\biggr|
> C_{+}\xi
\label{eq:7-27-2}
\end{eqnarray}
for every $\xi > \xi_{+}$. Put $\xi \equiv \Re\zeta$ and $- \eta
\equiv \Im\zeta$\, so that $\eta > 0$. Then, having $\xi \geq
\eta$, we get $2\xi^{2} - (\xi^{2} + \eta^{2}) = \xi^{2} -
\eta^{2} \geq 0$. Hence by (\ref{eq:7-27-1}) and (\ref{eq:7-27-2})
we obtain
$$\frac{C_{\pm}}{\sqrt{2}} \leq \frac{C_{\pm}\xi}{\sqrt{\xi^{2} +
\eta^{2}}} \leq \frac{|D_{\pm}\left(x,\kappa,\zeta + u^{-1}\left(
x\right)\right)|}{|\zeta|}$$
for $\zeta = \xi - i\eta$ with $\xi \geq \max\left(\xi_{\pm} ,
\eta\right)$. If $\zeta \in {\bf D}_{y, \theta_{0}}$ with
$\Re\zeta > \max\left(\xi_{\pm} , |\Im\zeta|\right)$, we have
\begin{eqnarray}
\frac{C_{\pm}}{\sqrt{2}}|\zeta| \leq \biggl|D_{\pm}\left(x ,
\kappa, \zeta + u^{-1}\left( x\right)\right)\biggr|\,.
\label{eq:7-27-3}
\end{eqnarray}
Using then (\ref{eq:*g}), {\em (a12)}, and (\ref{eq:7-27-3}), we
arrive at
$$|g_{x}\left( \zeta\right)| \leq
2\frac{\kappa^{2}\varrho(x)}{C_{+}C_{-}} C_{y}|\zeta|^{- (2 +
q_{y})}$$
for sufficiently large $|\zeta|$ with $\zeta \in {\bf D}_{\nu,
\theta_{0}}$. \quad \QED \vspace{2mm}

Next we set for any $x \in I_1$
\begin{eqnarray}
d_{\nu}^{x} \equiv x - \left(\nu + u^{-1}\left( x\right)\right) -
\kappa^{2}\varrho(x) \int_{\nu}^{\infty}\frac{v\left( u^{-1}\left(
x\right) , z\right)}{z - \nu}\, dz
\,.
\label{eq:7-28-1b}
\end{eqnarray}
\begin{remark}
{\rm Recall that by {\em (a5)} $d_{\nu}^{x}$
is positive for sufficiently small $|\kappa|$.}
\end{remark}

Let us finally state the last assumption:
\begin{description}
\item{\em (a13)} Given $x \in I_1$, there are
constants $A_{\nu,x} \ne 0$ and $p_{\nu,x} \geq 0$ such that
$$\mathop{\lim_{\zeta\to 0}}_{\zeta \in {\bf D}_{0,\theta_{0}}}
\frac{v\left( u^{-1}\left( x\right) , \zeta + \nu\right)}{
\zeta^{p_{\nu,x}}} = A_{\nu,x}\,.$$
\end{description}
We set
\begin{eqnarray}
\nonumber \kappa_{\nu,x}^{2} &:=& \frac{x - \nu - u^{-1}\left(
x\right)}{ \varrho(x)I\left( u^{-1}\left( x\right) , \nu +
u^{-1}\left( x\right) \right)} \\ \nonumber &{}& \qquad \\ &=&
\left( x - \nu - u^{-1}\left( x\right) \right) \left\{ \varrho(x)
\int_{\nu}^{\infty} \frac{v\left( u^{-1}\left( x\right) ,
z\right)}{ z - \nu}dz \right\}^{-1}\,.
\end{eqnarray}
By {\em (a5)}, this quantity satisfies
$\kappa_{\nu,x}^{2} > 0$
and $d^{x}_\nu\not= 0$ for $\kappa^2\not=\kappa_{\nu,x}^2$.
\begin{lemma}
\label{dlimit} Assume (a1)--(a3), (a7) and (a10). Then
\begin{equation}
\label{Dflim}
\lim_{\xi\to\nu^+,\,\eta\to 0^+}
D_{\pm}(x,\kappa,\xi-i\eta+u^{-1}(x))=d^x_\nu\, .
\end{equation}
\end{lemma}
{\sl Proof:} By (\ref{implf}), (\ref{Dminus}) and Lemma
\ref{analcont},
\begin{eqnarray*}
D_+(x,\kappa,\xi-i\eta+u^{-1}(x)) =
x-\xi-u^{-1}(x)+i\eta\quad\quad\quad\quad\quad\quad\quad \\
-\kappa^2\varrho(x) [\GG(u^{-1}(x),\xi-i\eta+u^{-1}(x))
+ 2i\pi v(u^{-1}(x),\xi-i\eta)]\, ,\\
D_-(x,\kappa,\xi-i\eta+u^{-1}(x)) =
x-\xi-u^{-1}(x)+i\eta\quad\quad\quad\quad\quad\quad\quad \\
-\kappa^2\varrho(x)\GG(u^{-1}(x),\xi-i\eta+u^{-1}(x))
\quad\quad\quad\quad\quad\quad\quad\quad\quad\quad\;\,\,
\end{eqnarray*}
for $\eta>0$, $\xi-i\eta\in\Omega$ which is the sufficient range
of variables as $\nu+u^{-1}(x)\in\Omega$ by {\em (a10)}. Under
assumption {\em (a10)}, there exists $A>0$ such that
$v(u^{-1}(x),\cdot)$ is holomorphic in the set $\{\zeta\in\C |
|\zeta-\nu|<2A\}$. Taking into account {\em (a7)} then
$$ \lim_{\xi\to\nu^+,\,\eta\to 0^+} v(u^{-1}(x),\xi-i\eta) =
v(u^{-1}(x),\nu)=0\, . $$
Let us write
\begin{eqnarray}
\nonumber
\GG(u^{-1}(x),\xi-i\eta+u^{-1}(x)) =
\int_\nu^{\nu+A}\frac{v(u^{-1}(x),z)-v(u^{-1}(x),\xi-i\eta)}
{z-\xi+i\eta}\, dz
\\
+\, v(u^{-1}(x),\xi-i\eta)\int_\nu^{\nu+A}\frac{dz}{z-\xi+i\eta}
\,+\,
\int_{A+\nu}^{\infty}\frac{v(u^{-1}(x),z)}{z-\xi+i\eta}\, dz\, .
\quad
\label{Gsep}
\end{eqnarray}
For the first and third integral, dominated convergence theorem
can be used giving (recall {\em (a7)})
$$ \int_\nu^\infty\frac{v(u^{-1}(x),z)}{z-\nu}\, dz $$
as the limit of their sum as $\xi\to\nu^+$, $\eta\to 0^+$. The
second integral
\begin{eqnarray*}
\int_\nu^{\nu+A}\frac{dz}{z-\xi+i\eta} &=&
\frac{1}{2}\ln\frac{(\nu+A-\xi)^2+\eta^2}{(\xi-\nu)^2+\eta^2}
\\&+&\,
i\left(\arctan\frac{\eta}{\nu+A-\xi}+\arctan\frac{\eta}{\xi-\nu} -
\pi\right)
\end{eqnarray*}
for $\nu<\xi<\nu+A$, $\eta>0$. As $|v(u^{-1}(x),\xi-i\eta)| \leq
c\sqrt{(\xi-\nu)^2+\eta^2}$ for $\xi-i\eta$ in a neighborhood of
$\nu$ with a suitable constant $c$, the limit of the second term
in (\ref{Gsep}) is zero. Now (\ref{Dflim}) is seen. \quad \QED
\vspace{2mm}
\begin{lemma}
\label{lem:A-g-d}
Let $0<|\kappa| < \delta_{2}$, $x\in I_1$ and
$d_{\nu}^{x} \not= 0$. Then
the function $g_{x}$ has no poles in $\left\{ \zeta \in
{\bf D}_{\nu, \theta_{0}}\, |\, |\zeta - \nu| <
\varepsilon_{0}\right\}$ with a constant $\varepsilon_{0} > 0$ and
the limit
\begin{eqnarray}
w_{\nu,x} &\!:=\!& \lim_{\zeta\to 0,\,\zeta\in {\bf D}_{0, \theta_{0}} }
\frac{g_{x}\left(\nu + \zeta\right)}{ \zeta^{p_{\nu,x}}} \\ &\!=\!&
\frac{\kappa^{2}\varrho(x)A_{\nu,x}}{ D_{+}\left(x,\kappa,\nu +
u^{-1}\left( x\right)\right) D_{-}\left(x,\kappa,\nu +
u^{-1}\left( x\right)\right)} \,=\,
\frac{\kappa^2\varrho(x)A_{\nu,x}}{{d_\nu^x}^2}\,. \nonumber
\end{eqnarray}
\end{lemma}
{\sl Proof:} The poles of $g_{x}(\zeta)$ come only from the zeroes
of $D_{\pm}\left(x,\kappa,\zeta + u^{-1}\left( x\right)\right)$.
If $d_\nu^x\not= 0$ then
\begin{equation}
\label{gxlim}
\lim_{\zeta\to\nu,\,\zeta\in {\bf D}_{\nu,\theta_0}} g_x(\zeta)
= 0
\end{equation}
by (\ref{eq:*g}) and Lemma \ref{dlimit}. By Lemma
\ref{lem:A-g-m}, $g_x$ is meromorphic in ${\bf D}_{\nu,\theta_0}$
so its only possible singularities there are isolated poles; they
also do not accumulate at $\nu$ due to (\ref{gxlim}).
Thus $g_{x}(\zeta)$ has no poles in a small neighborhood of $\zeta
= \nu$ in ${\bf D}_{\nu,\theta_{0}}$. By {\em (a13)}, we therefore
have
$$g_{x}\left(\nu + \zeta\right) \mathop{\sim}_{\zeta\to 0}
\frac{\kappa^{2}\varrho(x)A_{\nu,x}}
{{d_\nu^x}^2} \zeta^{p_{\nu,x}}\,.
\quad \QED $$
\vspace{2mm}

Now we can formulate the main theorem of this section:
\begin{theorem}
\label{thm:long-time-behavior} Assume (a1)--(a7), (a10)--(a13).
Then for every $x \in  I_{1}$ and $\kappa \in  \R$ satisfying
$\varrho(x)>0$, $d_{\nu}^{x} \ne 0$, $0<|\kappa| < \delta_{2}$ we
have the following asymptotic behaviour:
$${\cal U}(t , x) \mathop{\sim}_{t\to\infty}
w_{\nu,x}e^{-i[\nu+u^{-1}(x)]t}e^{- i\pi\left( p_{\nu,x} + 1\right)/2}
\Gamma\left( p_{\nu,x} + 1\right) t^{- (p_{\nu,x} + 1)},$$
where $\Gamma$ is the gamma function.
\end{theorem}
{\sl Proof:} It is sufficient to apply \cite[Theorem 2.1(b)]{Ar3}
to (\ref{eq:*u}) with the help of Lemmas \ref{lem:A-g-m}-\ref{lem:A-g-d}
and we obtain the desired result. \quad \QED \vspace{2mm}


\subsection*{Acknowledgment}

The research has been partially supported by GA ASCR and Czech
Ministry of Education under the contracts A1048801 and ME170. M.H.
was supported by Grant-in-Aid 11740109 for Encouragement of Young
Scientists from Japan Society for the Promotion of Science.



\begin{thebibliography}{99}
 %
\bibitem{Friedrichs}
K.O.~Friedrichs: ``On the perturbation of continuous spectra'',
{\em Commun. (Pure and) Appl. Math.} {\bf 1}, 361-406 (1948).
\vspace{-1.8ex}
 %
\bibitem{AMKG}
H.~Araki, Y.~Munakata, M.~Kawaguchi, T.~Goto: ``Quantum field
theory of unstable particles'', {\em Progr. Theor. Phys.} {\bf
17}, 419-442 (1957). \vspace{-1.8ex}
 %
\bibitem{DE}
J.~Dittrich, P.~Exner: ``A non-relativistic model of two-particle
decay I-IV'', {\em Czech. J. Phys.} {\bf B37}, 503-515, 1028-1034
(1987); {\bf B38}, 591-610 (1988); {\bf B39}, 121-138 (1989).
\vspace{-1.8ex}
 %
\bibitem{Ho1}
J.~Howland: ``Puiseux series for resonances at an embedded
eigenvalue'', {\em Pacific J. Math.} {\bf 55}, 157-176 (1974).
\vspace{-1.8ex}
 %
\bibitem{Ho2}
J.~Howland: ``The Livsic matrix in perturbation theory'', {\em J.
Math. Anal. Appl.} {\bf 50}, 415-437 (1975). \vspace{-1.8ex}
 %
\bibitem{BD}
H.~Baumg\"artel, M.~Demuth: ``Perturbation of unstable eigenvalues
of finite multiplicity'', {\em J. Funct. Anal.} {\bf 22}, 187-203
(1976). \vspace{-1.8ex}
 %
\bibitem{BDW}
H.~Baumg\"artel, M.~Demuth, M.~Wollenberg: ``On the equality of
resonances (poles of the scattering amplitude) and virtual
poles'', {\em Math. Nachr.} {\bf 86}, 167-174 (1978).
\vspace{-1.8ex}
 %
\bibitem{AG}
J.~Aguilar, J.-M.~Combes: ``A class of analytic perturbations for
one-body Schr\"odinger Hamiltonians'', {\em Commun. Math. Phys.}
{\bf 22}, 269-279 (1971). \vspace{-1.8ex}
 %
\bibitem{GWT}
H.~Grabert, U.~Weiss, P.~Talkner: ``Quantum theory of the damped
harmonic oscillator'', {\em Z. Phys.} {\bf B55}, 87-94 (1984).
\vspace{-1.8ex}
 %
\bibitem{APSJ}
C.~Aslangul, N.~Pitter, D.~Saint-James: ``Time behavior of the
cor\-relation functions in a simple dissipative model'', {\em J.
Stat. Phys.} {\bf 40}, 167-189 (1985). \vspace{-1.8ex}
 %
\bibitem{JIG}
R.~Jung, G.-L.~Ingold, H.~Grabert: ``Long-time tails in quantum
Brownian motion'', {\em Phys. Rev.} {\bf A32}, 2510-2512 (1985).
\vspace{-1.8ex}
 %
\bibitem{BM}
E.~Braun, P.A.~Mello: ``The correlation function for a quantum
oscillator in a low-temperature heat bath'', {\em Physica} {\bf
A143}, 547-567 (1987). \vspace{-1.8ex}
 %
\bibitem{Ar3}
A.~Arai: ``Long-time behavior of two-point functions of a quantum
harmonic oscillator interacting with bosons'', {\em J. Math.
Phys.} {\bf 30}, 1277-1288 (1989). \vspace{-1.8ex}
 %
\bibitem{Hi}
M.~Hirokawa: ``An inverse problem in quantum field theory and
canonical correlation functions. An application of a solvable
model called the rotating wave approximation'', {\em J. Math. Soc.
Japan} {\bf 51}, 337-369 (1999).\vspace{-1.8ex}
 %
\bibitem{Ar2}
A.~Arai: ``Spectral analysis of a quantum harmonic oscillator
coupled to infinitely many scalar bosons'', {\em J. Math. Anal.
Appl.} {\bf 140}, 270-288 (1989). \vspace{-1.8ex}
 %
\bibitem{KPPP}
E.~Karpov, I.~Prigogine, T.~Petrosky, G.~Pronko: ``Friedrichs
model with virtual transitions. Exact solution and indirect
spectroscopy'', {\em J. Math. Phys.} {\bf 41}, 118-131 (2000).
\vspace{-1.8ex}
 %
\bibitem{BFS1}
V.~Bach, J.~Fr\"{o}hlich, I.~M.~Sigal: ``Renormalization group
analysis of spectral problems in quantum field theory'', {\em Adv.
Math.} {\bf 137}, 205-298 (1998). \vspace{-1.8ex}
 %
\bibitem{BFS2}
V.~Bach, J.~Fr\"{o}hlich, I.~M.~Sigal: ``Return to equilibrium'',
{\em J. Math. Phys.} {\bf 41}, 3895-4060 (2000). \vspace{-1.8ex}
 %
\bibitem{BG}
F.~Barra, P.~Gaspard: ``Scattering in periodic systems: from
resonances to band structure'', {\em J. Phys.} {\bf A32},
3357-3375 (1999). \vspace{-1.8ex}
 %
\bibitem{RS1}
M.~Reed, B.~Simon: {\em Methods of Modern Mathematical Physics. I.
Functional Analysis}, Academic Press, New York, 1972.
\vspace{-1.8ex}
 %
\bibitem{JarDII}
V.~Jarn\'\i k: {\em Differential Calculus II}, Academia, Prague,
1956 (in Czech).\vspace{-1.8ex}
 %
\bibitem{Cerny}
I.~\v{C}ern\'{y}: {\em Foundations of Analysis in the Complex
Domain}, Academia, Prague, 1967 (in Czech). \vspace{-1.8ex}
 %
\bibitem{Rudin}
W.~Rudin: {\em Real and Complex Analysis}, McGraw-Hill, New York,
1974.
 %
\end{thebibliography}
\end{document}